%% file: TeV_paper.tex
\documentclass[structabstract,twocolumn,twoside,a4paper]{aa}

\usepackage{geometry}
\usepackage[english]{babel}
\usepackage{natbib}
\usepackage{graphicx}
\usepackage[chatter]{rotating}
\usepackage{epsfig}
\usepackage{scrextend}

\usepackage{lineno}             
\usepackage{amssymb}        
\usepackage{longtable}          
\usepackage{lscape}
\usepackage[usenames,dvipsnames]{color}
\usepackage[normalem]{ulem}
\usepackage{amsmath}
\usepackage{multirow}

\usepackage{booktabs}

\def\gsim{\mathrel{\rlap{\lower4pt\hbox{\hskip1pt$\sim$}}
    \raise1pt\hbox{$>$}}}                
\bibpunct{(}{)}{;}{a}{}{,} 

\begin{document}

\title{Search of MeV-GeV counterparts of TeV sources with AGILE in pointing mode}


%
%
%
\include{author-list-aa}

\date{}

\abstract{
Known TeV sources detected by major 
\v{C}erenkov telescopes are investigated to identify possible 
MeV-GeV $\gamma$--ray counterparts.
}
{A systematic study of the known sources in the web--based \textit{TeVCat} 
Catalog has been performed to search for possible $\gamma$--ray counterpart
on the AGILE data collected during the first period of operations
in observing pointing mode.
}
{For each TeV source, a search for a possible $\gamma$--ray 
counterpart based on a multi-source Maximum Likelihood 
algorithm is performed on the AGILE data taken 
with the GRID instrument from July 2007 to October 2009.}
{
In case of high-significance detection, the average $\gamma$--ray flux 
is estimated. For the cases of low-significance detection 
the 95\% Confidence Level (C.L.) flux upper limit is given.
52 TeV sources out of 152 
(corresponding to $\sim 34\% $ of the analysed sample) show a significant
excess in the AGILE data covering the pointing observation period.
}
{
This analysis found 26 new AGILE sources with respect to the AGILE reference catalogs, 
 15 of which are Galactic, 7 are extragalactic and 4 are 
unidentified. Detailed tables with all available information on the 
analysed sources are presented.
}


\titlerunning{TeV sources detected by AGILE}

\maketitle

\section{Introduction}
\label{sec:INTRODUZIONE}
In the last years, the number of identified TeV sources 
has increased up to more than 100, thanks to 
the observations made by the new generation of ground-based 
\v{C}erenkov telescopes HESS~\citep{Hinton:2004eu}, 
MAGIC~\citep{Ferenc:2005fd} and VERITAS~\citep{Holder:2006gi}. 
These sources mainly belong to five classes: Active 
Galactic Nuclei (AGN), Supernova Remnants (SNR),
Pulsar Wind Nebulae (PWN), X-rays binary systems (XRB), and Pulsars
(PSR).
More than 80 TeV sources are Galactic and a significant 
fraction of them ($ \gsim 20 \%$) 
does not show any evident counterpart and remains unidentified (UNID).

Multi--wavelength deep observations of the regions near 
the TeV sources are needed to identify the possible 
counterparts of the UNID, as well as to understand the 
emission mechanisms of the TeV $\gamma$--rays.

Following previous studies on positional and spectral connection 
of GeV to TeV $\gamma$--ray sources  
done for EGRET and Fermi LAT \citep{funk1, funk2, abdoagn, aceropwn}, 
this paper reports the results of the search for $\gamma$--ray 
emission from known and unidentified TeV sources, using the data collected 
by AGILE in pointing mode in the energy range above 100 MeV.

This search is particularly relevant, as demonstrated by 
the previous studies, due to to fact that two 
adjacent energy ranges probe different regions of the source spectra.
Preliminary results have been previously 
presented \citep{RICAP09,FERMISYMPOSIUM}.

\begin{figure*}[ht]
\begin{center}
\includegraphics[width=0.8\textwidth]{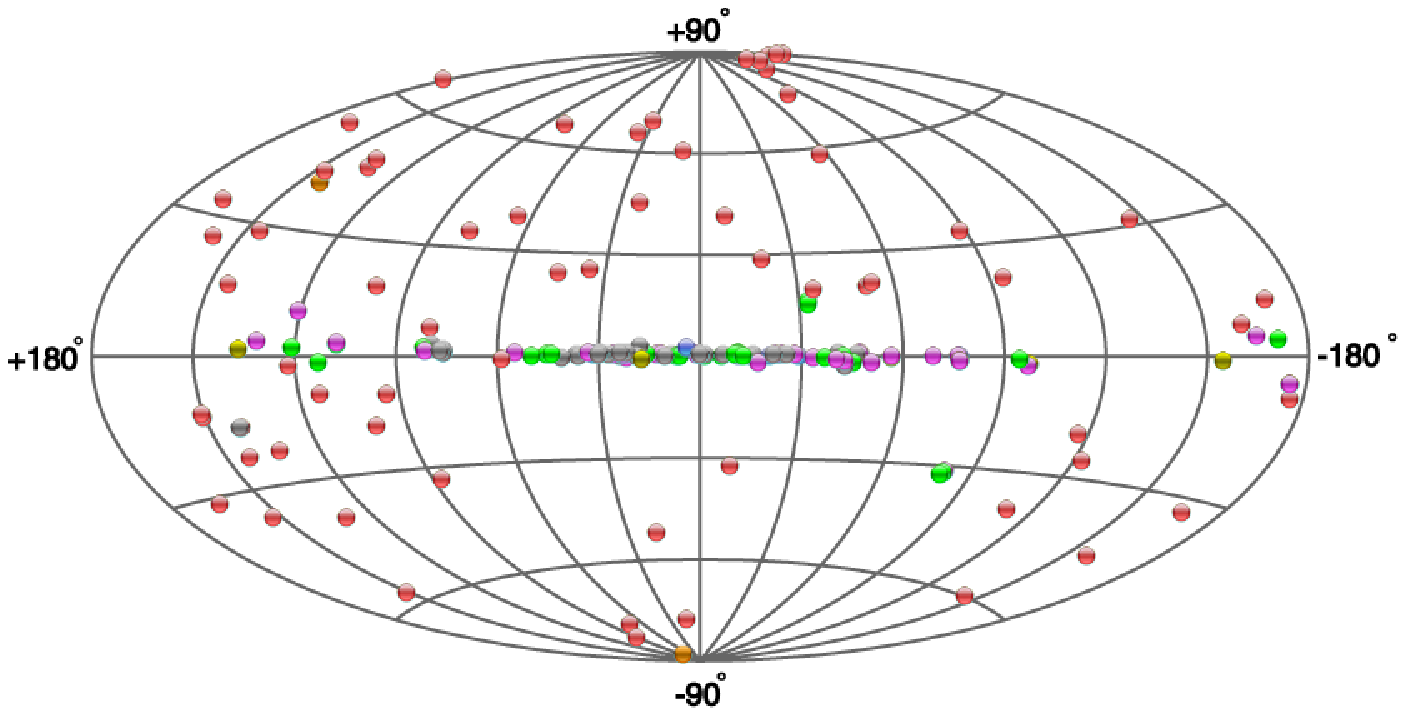}
\end{center}
\includegraphics[width=0.3\textwidth]{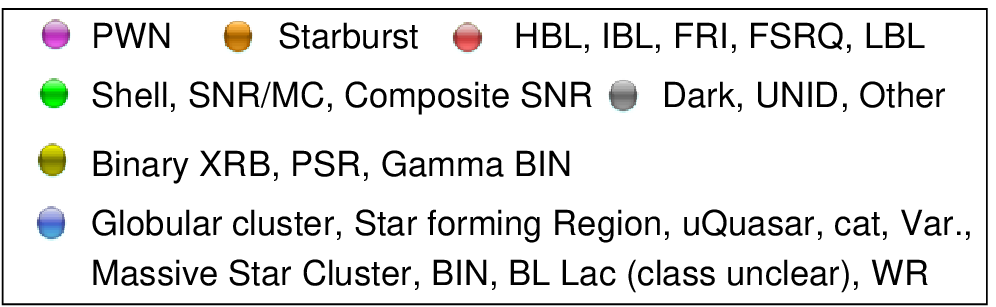}
\caption{The Aitoff projection in Galactic coordinates of the 
TeV source positions, as extracted from the on-line {\it TeVCat}
({\it Default Catalog} and {\it Newly Announced} samples, June 2015)~\citep{TEVCAT}.
}
\label{fig:tevcat}
\end{figure*}

\section{The TeV source Catalog}
\label{sec:SOURCES}

The analysis described in this paper has been applied to 
a reference sample of TeV sources extracted from 
the on-line {\it TeVCat} Catalog\footnote{http://tevcat.uchicago.edu/}~\citep{TEVCAT}.
This on-line catalog
is continuously updated 
with new sources detected by the TeV experiments, and for each source
it provides many parameters such as coordinates, source
type, flux and estimated distance (when available).


At the time of writing (June 2015), the {\it TeVCat} 
Catalog contains a total of 183 TeV sources: 129 of those are 
flagged as {\it Default Catalog} and have been published 
on refereed journals, 32 are flagged as 
{\it Newly Announced} (see also Fig.~\ref{fig:tevcat}),
10 are flagged as {\it Other Sources} and 
12 are flagged as {\it Sources Candidates}. 
The analysis described here was performed on a subset of 152 
TeV sources, both Galactic and extragalactic, consisting of 
120 sources of the {\it Default Catalog}, plus 32
{\it Newly Announced} sources. 
The following criteria were adopted to define the input sample.
Three {\it extended regions}, 
which already include a compact TeV source, were 
excluded: the Galactic Centre Ridge (including HESS J1745-290 and SNR G 0.9+0.1), Boomerang PWN 
(including SNR G 106.3+2.7) and Milagro Diffuse (including MGRO J2019+37). 
The TeV sources SN 1006 SW and NE, HESS J1018-589 A and B 
as well as HESS J1800-240 A and B, were considered as single 
candidate $\gamma$--ray sources located, respectively, at the centre of 
the SNR 1006 shell, and at the centroid positions of HESS J1018-589 and 
HESS J1800-240. 
The two TeV sources ARGO J2031+4157 and MGRO J2031+41 were not included
since they are both associated with TeV J2032+415 in the {\it TeVCat}.
Moreover, the detection of the pulsed emission at TeV energies 
from the Crab and Vela Pulsar has been not considered in the sample
since the timing analysis of pulsars is out of the scope of this paper.

Each sky position and extension of the TeV sources 
in the selected {\it TeVCat} sample has been carefully reviewed 
using published data. A new interactive web page of the 
catalog of TeV sources, including this 
coordinate revision and giving public access to light curves and spectra, 
is now available at the ASI Science Data Center 
(ASDC)~\citep{ASDC-TeV}\footnote{http://www.asdc.asi.it/tgevcat/}.

When available, the best-fit position of the TeV excess 
has been used as starting input position for the AGILE data
analysis. Otherwise, the position of the optical/radio known
counterpart has been used. 
The error region on each TeV source position has been calculated 
by summing quadratically the statistical uncertainties on the 
position coordinates 
obtained from the 2D-Gaussian fit of the TeV excess, and the systematic 
uncertainties on the instrument pointing (when available 
in the literature).
%

\section{The AGILE satellite}
\label{sec:Instrument}
AGILE~\citep{MISSION} is an Italian Space Agency (ASI)
small scientific mission for high-energy astrophysics launched
on April 23, 2007 from the Indian base of Sriharikota 
in an equatorial orbit optimised for low particle 
background, with a very small inclination angle 
($ \sim $2.5$^\circ$) and initial altitude of about 550 km. \par
The analysis has been performed using
the data collected by the main AGILE instrument, 
the {\it Gamma--Ray Imaging Detector} (GRID).   
The AGILE-GRID is sensitive in the energy range 30 MeV -- 50 GeV, 
and it consists of a silicon-tungsten tracker, a caesium iodide 
mini-calorimeter and an anticoincidence system made 
of segmented plastic scintillators.
The use of the silicon strip technology allows to have 
good performance for the $\gamma$--ray GRID imager,
approximately a small cube of $\sim$ 60 cm size, 
which achieves an effective area of the order of 500~cm$^2$ 
at several hundreds MeV, an angular resolution (at 68\% containment radius)
of about 4.3$^\circ$ at 100 MeV, decreasing below 1$^\circ$ 
for energies above 1 GeV~\citep{chenaa}, an unprecedentedly large 
field of view (FOV) of about $\sim 2.5$ sr, as well as 
accurate timing, positional and attitude information (source location
accuracy $5^\prime$ -- $10^\prime$ for intense sources with S/N $ \gsim 10$).

\section{AGILE data set}
\label{sec:DATA}
During its first period of data taking (about two years)
the AGILE satellite was operated in ``pointing observing mode'',
and the corresponding AGILE data are divided in 
{\it Observation Blocks} (OBs). Each AGILE OBs consists 
of long exposures, which mostly range from a few days 
to about thirty  
days, with the pointing direction 
drifting $\sim 1^\circ$ per day with respect to the 
initial boresight direction to match solar panel 
illumination constraints. 
\par
\begin{figure*}[htb]
\begin{center}
\includegraphics[width=0.8\textwidth]{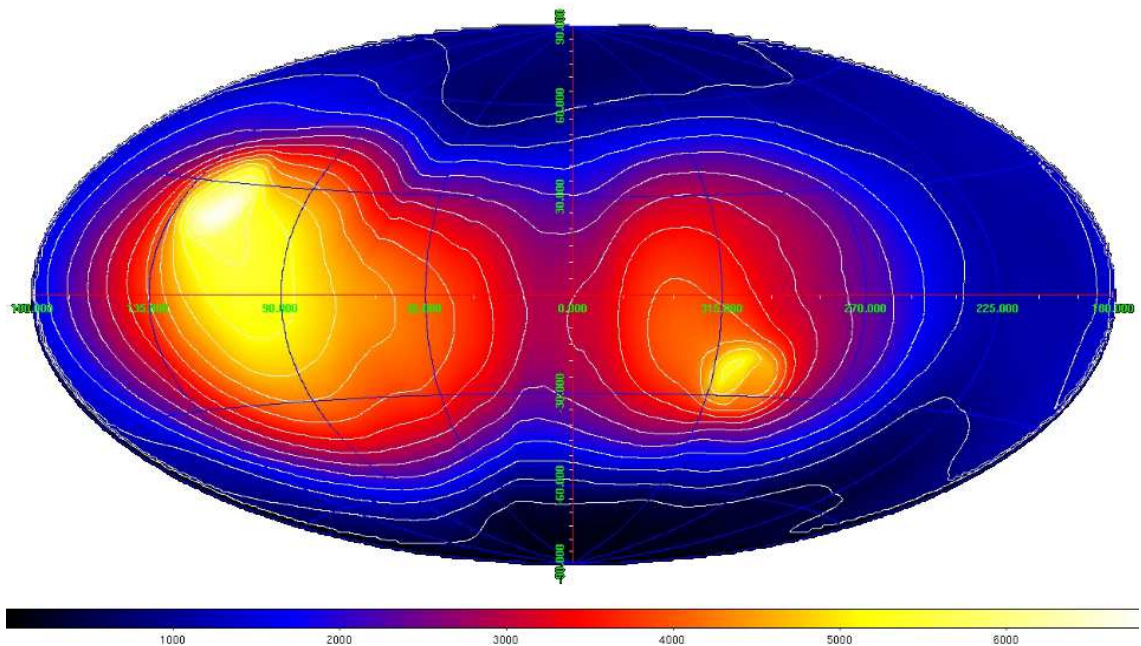}
\end{center}
\caption{Total AGILE exposure map during the first $\sim$ 
two years of operations (July 2007 -- October 2009). The 
exposure values are expressed in [$\mathrm{cm^{2} ~Ms}$].
The mean, maximum and minimum exposures attained correspond to values of about 
2250, 6800 and 60 $\mathrm{cm^{2} ~Ms}$,
respectively.
} 
\label{fig:exposure}
\end{figure*}

The analysed AGILE data set covers the period from July 9, 2007 
(beginning of the science verification phase)
to October 18, 2009, corresponding to 96 OBs (not 
including the first 5 OBs of the Commissioning). 
During this period, following the main scientific 
program of the AGILE mission, the satellite was mainly 
pointed to observe two regions near the Galactic
Plane, around $l = 90^\circ$ and $l = 270^\circ$ longitude values,
as shown by the total exposure map in Fig.~\ref{fig:exposure}.
Clearly, this observation strategy mainly focused on 
the Galactic plane was not optimal for detecting 
extragalactic TeV sources.

\section{Data analysis procedure}
\label{sec:ANALYSIS}
An iterative automated procedure has been developed to analyse 
the entire pointing AGILE data set described in Sec.~\ref{sec:DATA}, 
searching for possible $\gamma$--ray excesses correlated to the TeV sources. 
For each TeV source in the reference sample defined 
in Sec.~\ref{sec:SOURCES}, this procedure 
is divided in two main parts.
\par
\subsection{Maps creation}
\label{sec:maps}
The first 
part consists in the creation of 
the maps of counts, exposure 
and diffuse $\gamma$--ray background based on the model described in \citep{Giuliani:2004saif} updated together with the new GRID Calibrations, centred at the TeV 
source position, over the full AGILE pointed observation period.
These maps are in Galactic coordinates ({\it l, b}), ARC projection,  
with size of $40^\circ\times 40^\circ$ divided in bins 
of $0.1^\circ\times 0.1^\circ$, and they have been produced 
using the latest official AGILE scientific analysis software~\citep{AGILESW}
-- available at ASDC -- with the following parameters:

\begin{itemize}
  \setlength{\itemsep}{0pt}
  \item {\it data archive:} {\sf ASDCSTDe}
  \item {\it initial time:} {\sf MJD (TT) = 54290.5 (MET=111067134 s)\footnote{MET is the AGILE Mission Elapsed Time in seconds since 2004.0 UTC.}}
  \item {\it final time:} {\sf MJD (TT) = 55122.5 (MET=182951934 s)}
  \item {\it energy range:} {\sf 100 MeV $\div$ 50 GeV}
  \item {\it software release:} {\sf BUILD21}
  \item {\it event filter:} {\sf FM3.119}
  \item {\it response matrices:} {\sf I0023}
\end{itemize}

The {\it event filter} is the algorithm that processes the GRID raw data
and reconstructs the energy and direction
of the incident $\gamma$--ray in the GRID reference system.
The event filter used in this analysis ({\sf FM3.119})
represents the most updated reconstruction
algorithm, which provides a good trade-off between FoV,
effective area and background rejection \citep{AGILESW,chenaa}.
Maps were generated for energies E $>$ 100~MeV,  
including all events collected up to $60^{\circ}$ off-axis.
The South Atlantic Anomaly data were excluded, and
to eliminate the Earth albedo contamination,
events with reconstructed directions with respect to
the satellite-Earth vector smaller than $85^{\circ}$ were also
rejected.

\subsection{Source detection}
\label{sec:detection}
The next part of the automatic procedure consists 
in verifying if around the input TeV position it 
is possible to detect the presence of a 
significant $\gamma$--ray source
and, if so, to try to locate the best position
of the $\gamma$--ray excess.

The source detection is performed by means of a 
{\it multi-source} Maximum Likelihood Estimator
(MLE) algorithm that estimates the photon counts and the 
position of the TeV source, the expected contribution
given by the background components (modelled as a superposition
of a Galactic diffuse emission background and an isotropic component),
all the known AGILE $\gamma$--ray sources within
the region of analysis, taking also into account the
instrument response function.

In particular, the MLE analysis was performed taking into 
account all the known $\gamma$--ray sources detected by AGILE
at the time of the analysis,  
consisting of a set of 65 sources, obtained 
by combining the 54 sources of the 
updated list of AGILE bright sources~\citep[1AGLR;] [] {AGILEVAR},
plus 11 sources not bright enough to be detected over the short OB time scales of the 1AGLR analysis:
i.e. eight 1AGL sources from the first AGILE high confidence Catalog ~\citep{AGILECAT}
and two AGL sources in the Carina region (see also Tab.~2 and Tab.~3 
of 1AGLR paper), plus one additional AGL source from a detailed analysis of the 
Cygnus region \citep{Bulgarelli:2012cyg}.
All AGILE sources are assumed to be point-like with simple power-law spectra.

With the exception of bright sources (significance $> 5$),
AGILE data analysis may not be spectrally resolved due to
low statistics, and in general a standard fixed spectral
spectral index value of -2.1 is adopted for the initial steps of the ML analysis.
This assumption is motivated by the known spectral properties
of the majority of the $\gamma$--ray sources in the AGILE energy range, 
except for few sources as described in the 1AGLR.
Timing analysis of pulsars was not performed in this paper, 
and the $\gamma$--ray emission detected in the search of counterparts to the TeV 
emission from PWN is in general due to the average (pulsar + nebula) $\gamma$--ray flux values.

\par
The significance of a source detection is evaluated by 
the square root of the {\it test statistic} $TS$, defined as
\begin{equation}
TS = -2 \; \log\left(\frac{\mathcal{L}_0}{\mathcal{L}_1}\right)
\label{eq:ts}
\end{equation}
where $\mathcal{L}_0/\mathcal{L}_1$   
is the ratio between the maximum likelihood $\mathcal{L}_0$ of the null 
hypothesis and the likelihood 
$\mathcal{L}_1$ of the alternative hypothesis (presence 
of a point--like source under evaluation)~\citep{Mattox:1996zz,Bulgarelli:2012he}. 
For a large enough number of counts ($N \gsim 20$), $TS$ is 
expected to behave as $\chi^{2}_{1}$ in the null 
hypothesis, and the significance of a source detection 
is given by $\sqrt{TS}$.

The source detection algorithm is very flexible and can be used 
with a variety of parameters and options that allow to 
refine the process of source detection and location.
For example, both the position and the flux of the analysed 
source can be considered fixed or variable (starting from 
a defined initial value), as well as the coefficients of 
the Galactic background (diffuse emission and isotropic 
component) may in turn be kept fixed or treated as variable.  

In order to get the best result for the position and 
flux estimation of the analysed sources, an iterative 
MLE analysis is performed, divided into the following steps:

\begin{itemize}

\setlength{\itemsep}{6pt}

\item {\bf step 1}: the aim of the first step is
both to find an excess of $\gamma$--rays around the
input TeV source position in the AGILE data,
and to estimate by the minimisation algorithm the best values of 
the $\gamma$--ray background model parameters
(Galactic and extragalactic isotropic contributions).
The MLE analysis is performed allowing the 
source coordinates to vary within a distance $\le 1^\circ$
from the input TeV source position,
taking into account all known AGILE sources
in the region of analysis with a radius of $10^\circ$;


\item {\bf step 2}: in this step the $\gamma$--ray excess
position is refined by fixing the Galactic background 
coefficients to the best values obtained from step 1;

\item {\bf step 3}: this is the final step to
estimate the flux and significance $\sqrt{(TS)_3}$
 of the $\gamma$--ray source at the optimised position 
resulting from step 2, using updated Galactic background 
coefficients at the new position;

\end{itemize}

\begin{itemize}
\item {\bf step 4}: this step gives directly
an estimation of the flux and significance
$\sqrt{(TS)_4}$ of the $\gamma$--ray source 
assuming a fixed position coincident with that of
the input TeV source. In this case the diffuse background coefficients 
are estimated at the original input position.

\end{itemize}

{\em Step 4}  represents the standard method used in literature to 
verify the significance of known sources
at input positions already known in other wavelengths. However, 
especially in the analysis of crowded regions of the Galactic plane,
it is important also to search for a possible optimised position 
of the $\gamma$--ray excess, 
still compatible with the TeV source spatial association. This 
search is performed in {\em step 1 -- step 3}.
\par

The possible shift in the $\gamma$--ray excess position 
may be due to several factors: 
\begin{itemize}

\item the rather poor angular resolution
of the order of a few degrees in the MeV--GeV energy range\footnote{
About $0.7^{\circ}$ PSF HWHM at 400 MeV, corresponding to
the AGILE effective area peak values \citep{Sabatini:2015nqa}.};
 
\item the AGILE reference catalogs positioning errors 
(ranging from $\sim 0.1^{\circ}$ for very bright sources up to $0.7^{\circ}$  
for faint sources, at $95\%$ C. L.); 

\item extended TeV source (X);

\item the possible physical displacement between the TeV and the MeV--GeV 
emission regions. 
\end{itemize}


In this paper an AGILE {\it detection} is in general defined by the condition:
\begin{equation}
\sqrt{(TS)_4} \ge 4
\label{eq:cond2}
\end{equation}
which corresponds to a statistical significance 
of about 4 $\sigma$ at the input TeV source position.

The search for optimised positions of the $\gamma$--ray excess
gives results which
are considered reliable when 
the MLE analysis ({\em step 1 -- step 3})
converges well within the allowed searching distance from
the input position, and the significance of 
the detection is increased. 
In practice (see next section) when the following condition is satisfied:

\begin{equation}
\sqrt{(TS)_3} \ge 4 \text{ and } \text{ and } dist \lesssim 0.6^\circ
\label{eq:cond1}
\end{equation}

where $dist$ is the angular distance between the position of the input
TeV source and the candidate $\gamma$--ray source.\\

\section{Results}
\label{sec:RESULTS}

In Tab.~\ref{tab:all} the complete list of all the TeV 
sources considered in this analysis is reported, 
showing the following relevant source parameters:
\begin{itemize}
\item{{\sf ID}: identification number;
}
\item {{\sf TeV source}: TeV source name;
}
\item { {\sf (l,b)}: position of the TeV source in Galactic Coordinates;
}
\item {{\sf TeV Pos. Err.}: positional error of the input TeV source (derived as explained in Sec.~\ref{sec:SOURCES});
}
\item {{\sf Canonical name};
}
\item {{\sf Type}: type of TeV source counterpart, if already known from other wavelengths;
}
\item {$\sqrt{(TS)_4}$: estimate of the $\gamma$--ray statistical significance 
at {\em step 4};
}
\item {$\sqrt{(TS)_3}$: estimate of the $\gamma$--ray statistical significance 
at {\em step 3};
}
\item {{\sf Flux$_4$}: estimate of the $\gamma$--ray flux (E$>$100~MeV)  
and its 1$\sigma$ statistical error in units of $\mathrm{10^{-7}\; ph\; cm^{-2}\; s^{-1}}$,
for AGILE detected sources at the input TeV source position from {\em step 4};
}
\item {{\sf UL$_4$}: AGILE $\gamma$--ray flux upper limit at 95\% 
C.L. at the input TeV source position from {\em step 4};
}
\item {{\sf EXT}: TeV source extension.
}
\end{itemize}

\begin{figure*}[htb]
\begin{center}
\includegraphics[width=0.8\textwidth]{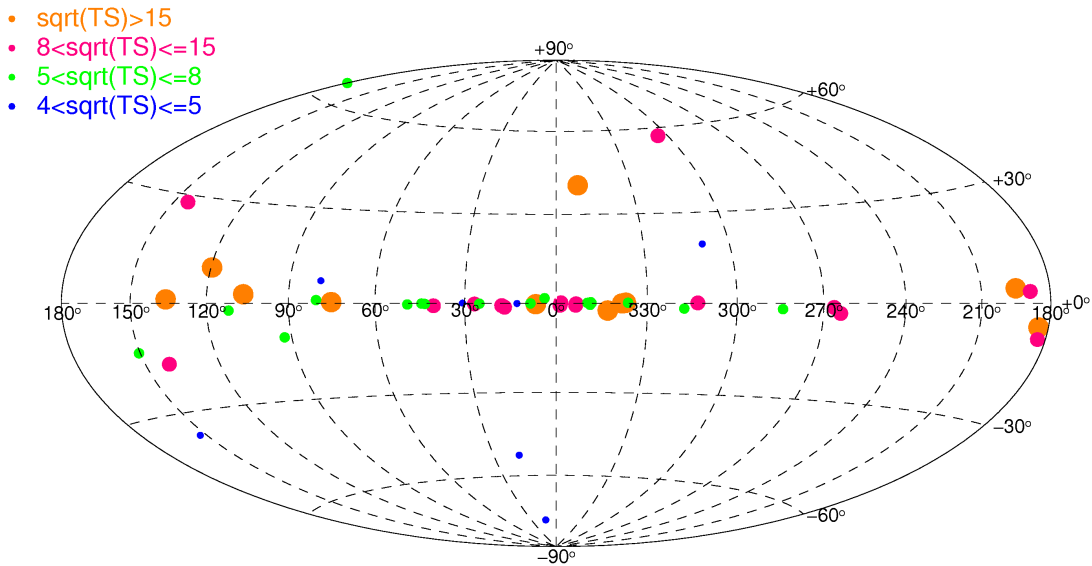}
\end{center}
\caption{Aitoff map in Galactic coordinates of all the 
detections according to the
criteria specified in the text,
corresponding to TeV sources showing a significant
excess in the AGILE data.
}
\label{fig:detected}
\end{figure*}

In Tab.~\ref{tab:all} the $\gamma$--ray sources detected
according to the criteria specified in
Sec.~\ref{sec:detection} are shown in bold.
For these sources the calculated flux value and the corresponding
error is given, while for the sources not satisfying the detection
requirement, the estimated 95\% C.L upper limit is reported.

In total, 52 TeV sources show a significant
excess in the AGILE data covering the pointed observation
period, corresponding to 34\% of the original sample.
The Aitoff map in Galactic coordinates of all the detections
is shown in Fig.~\ref{fig:detected}.

Tab.~\ref{tab:good_AGL} groups all the sources detected 
with AGILE in this work (shown in bold in the previous table), 
and it includes the following columns:

\begin{itemize}
\item{{\sf ID}: source identification number used in Tab.~\ref{tab:all}
}
\item{{\sf TeV Source}: TeV source name;
}
\item{$\sqrt{(TS)}$: estimate of the $\gamma$--ray source statistical significance
as result of the AGILE MLE analysis (upper part {\em step 3}, lower part {\em step 4};
}
\item{({\sf l,b}): optimised peak position of the AGILE excess in 
Galactic Coordinates (upper part);
}
\item{{\sf Error}: $\gamma$--ray source location error radius at 95\% C.L. from {\em step 3} (statistical error only);
}
\item{{\sf Flux}: estimate of the $\gamma$--ray flux (E$>$100~MeV) 
at the optimised peak position and its 1$\sigma$ statistical 
error in units of $\mathrm{10^{-7}\; ph\; cm^{-2}\; s^{-1}}$; 
}
\item{{\sf Dist}: distance of the $\gamma$--ray peak position from the input position of 
the TeV source;
}
\item{{\sf AGILE association}: 
already known AGILE source from the published 1AGL/1AGLR
catalogs~\citep{AGILECAT, AGILEVAR} within the error radius in the 7th column; 
}
\item{{\sf Fermi association}: known Fermi-LAT source(s) associated to the TeV source,
as described on 3FGL Catalog~\citep{3FGL}; 
}
\item{{\sf Analysis flag} (see below).
}
\end{itemize}

Tab.~\ref{tab:good_AGL} is split in two sections: 
the upper part reports the results of MLE analysis {\em step 3} 
for all the sources satisfying the condition in Eq.~(\ref{eq:cond1}).
The lower part includes the detected sources 
satisfying the condition in Eq.~(\ref{eq:cond2}) but not Eq.~(\ref{eq:cond1}),
for which the {\em step 3} automatic analysis is not reliable.
For this reason, the values of $\sqrt{TS}$ and flux shown in this part
of the table are those found at {\em step 4} at the input TeV positions. 
There are few exceptions to these criteria which are described in the 
table footnotes. For all the 19 sources in the lower section of Tab.~\ref{tab:good_AGL}, 
although the MLE analysis result at fixed input position is significant,
the tentative optimisation of the location of the $\gamma$--ray peak 
flux fails. In this cases the region of analysis may be not yet 
well modelled, and a refined MLE analysis should be performed after the 
release of new AGILE Catalog (Bulgarelli et al., in preparation).

As reported in the 8th column of Tab.~\ref{tab:good_AGL}, 
26 spatial associations 
of the detected TeV sources 
with already known AGILE sources
from the the 1AGL/1AGLR catalogs are found within the 95\% C.L. error radius (7th column)
~\citep{AGILECAT, AGILEVAR}.
Among these sources, 15 are 
Galactic, 6 are extragalactic and 5 are unassociated.
As reported in the 9th column, 46 spatial associations
with known Fermi sources from the 3FGL Catalog 
are found 
(note that some TeV sources have more than one 3FGL association).
Fermi counterparts officially 
associated in the 3FGL Catalog to the corresponding TeV 
source may have flag P (for point-like sources) or E (for 
extended sources). 

\begin{figure}[htb]
\begin{center}
\includegraphics[width=1.0\columnwidth]{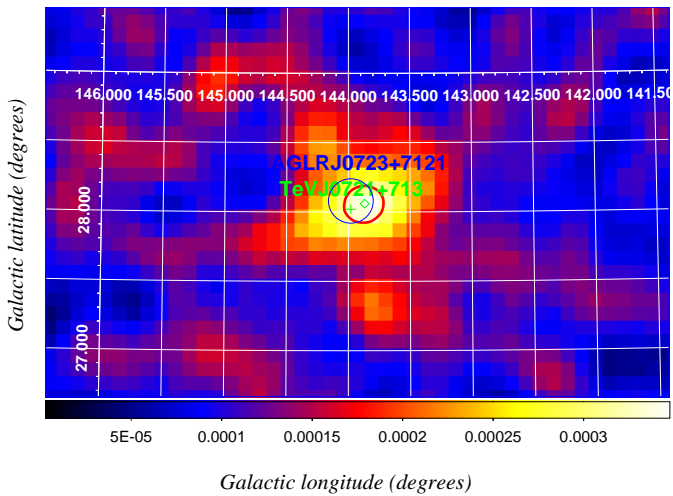}
\end{center}
\caption{Source detection in the position of TeV J0721+713. The
signal is detected with $\sqrt{(TS)_3}$ = 13.9, and the
localization algorithm gives the 95\% C.L. contour
(red curve) that entirely contains the TeV source
position (green cross). The AGILE counterpart 1AGLR J0723+7121
is also shown with its error radius (blue circle).
The picture shows the {\it intensity} map
(cm$^{-2}$s$^{-1}$bin$^{-1}$) in galactic coordinates,
with bin size $0.1^{\circ}\times0.1^{\circ}$
and Gaussian smoothing of three bins' radius.
}
\label{fig:Source_in}
\end{figure}

\begin{figure}[htb]
\begin{center}
\includegraphics[width=1.0\columnwidth]{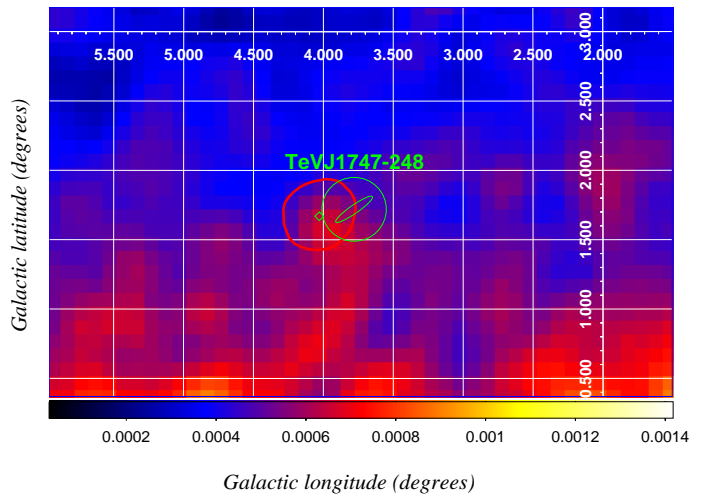}
\end{center}
\caption{Source detection in the position of TeV J1747-248. The 
signal is detected with $\sqrt{(TS)_3}$ = 5.9, and the 
localization algorithm gives the 95\% C.L. contour 
(red curve) which partially overlaps with the error circle and 
the extension of the TeV source (green circle and green ellipse).
The picture shows the {\it intensity} map
(cm$^{-2}$s$^{-1}$bin$^{-1}$) in galactic coordinates,
with bin size $0.1^{\circ}\times0.1^{\circ}$
and Gaussian smoothing of three bins' radius.
}
\label{fig:Source_overlap}
\end{figure}

\begin{figure}[htb]
\begin{center}
\includegraphics[width=1.0\columnwidth]{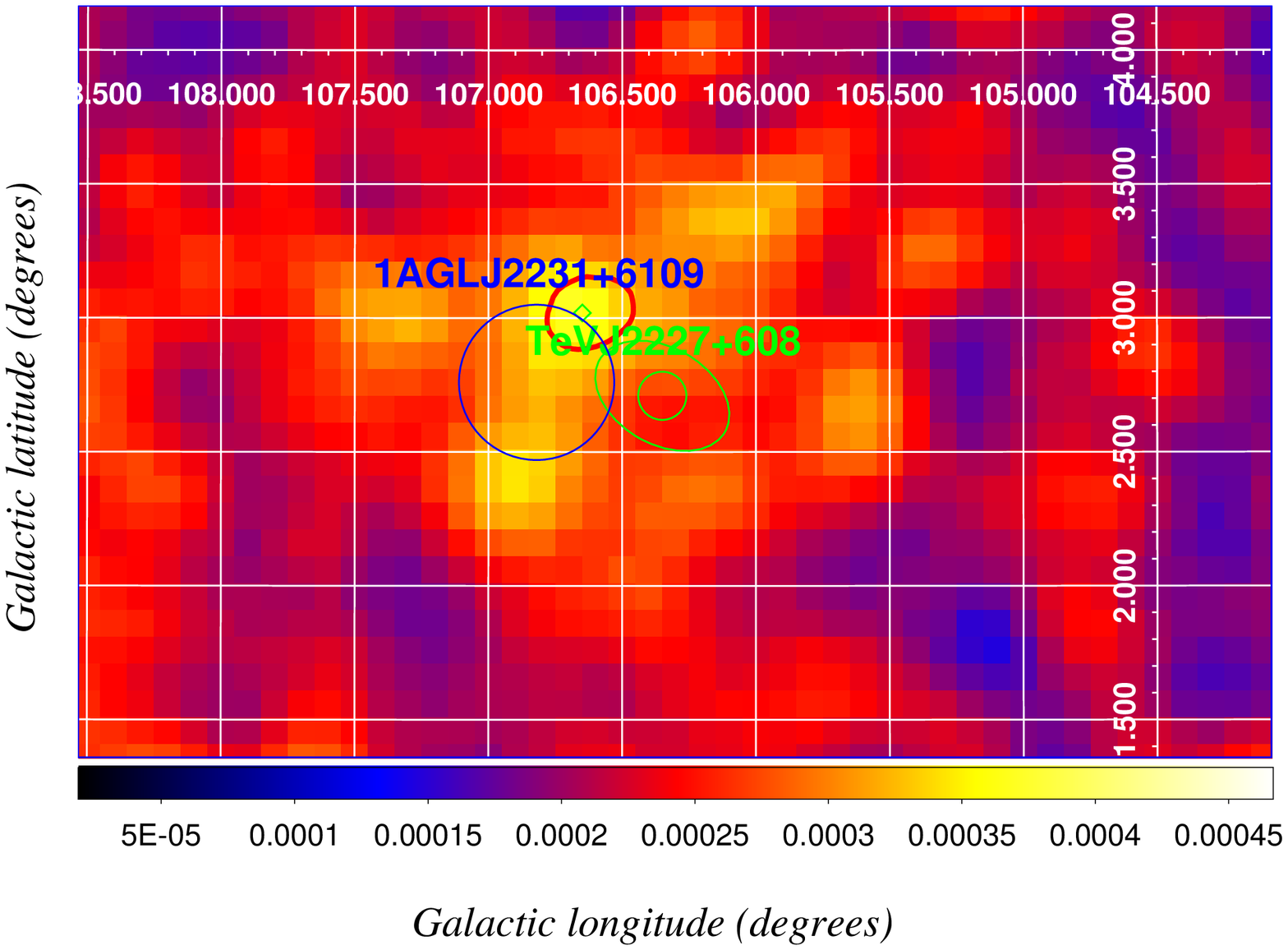}
\end{center}
\caption{Source detection in the position of TeV J2227+608. The 
signal is detected with $\sqrt{(TS)_3}$ = 16.7, and the localization 
algorithm gives the 95\% C.L. contour (red curve) which 
neither includes nor overlaps with the error circle and 
the extension of the TeV source (green circle and 
green ellipse). The AGILE counterpart 1AGL J2231+6109 is also 
shown with its error radius (blue circle).
The picture shows the {\it intensity} map
(cm$^{-2}$s$^{-1}$bin$^{-1}$) in galactic coordinates,
with bin size $0.1^{\circ}\times0.1^{\circ}$
and Gaussian smoothing of three bins' radius.
}
\label{fig:Source_external}
\end{figure}

Column 10th of Tab.~\ref{tab:good_AGL} reports an analysis flag 
assigned to the AGILE detection according to the position and 
extent of the source location contour:
\begin{itemize}
\item {\it IN} (Inside): the TeV source, including its 
extension (if any), is entirely within the AGILE 
contour (see an example in Fig.~\ref{fig:Source_in});
\item {\it O (Overlapping)}: the AGILE contour at the 
95\% C.L. overlaps with the error circle and/or the extension 
of the TeV source (see Fig.~\ref{fig:Source_overlap});
\item {\it E (External)}: the AGILE contour neither includes 
nor overlaps with the TeV source. Nevertheless, the 
AGILE peak position is within 0.6$^\circ$ from the 
TeV source position (see Fig.~\ref{fig:Source_external}).
\end{itemize}

\par
In this work 26 new AGILE sources are found
with respect to the AGILE reference catalogs, 
15 of which are Galactic, 7 are extragalactic and 4 are 
unidentified. A detailed statistics about the type of the 
detected source can be found in Tab.~\ref{tab:statistics}. 

There are 8 sources detected by AGILE in this analysis with 
no Fermi 3FGL official association:

\begin{itemize}
\item[*]{ID 88:  TeV J1634-472 (HESS J1634-472)}
\item[--]{ID 96:  TeV J1713-382 (CTB 37B)}
\item[--]{ID 103:  TeV J1729-345 (HESS J1729-345)}
\item[--]{ID 104:  TeV J1732-347 (HESS J1731-347)}
\item[--]{ID 105: TeV J1741-302 (HESS J1741-302)}
\item[*]{ID 116: TeV J1813-178 (HESS J1813-178)}
\item[--]{ID 133: TeV J1911+090 (W49B)}
\item[--]{ID 134: TeV J1912+101 (HESS J1912+101)}
\end{itemize}

two of which, indicated by asterisks, have counterparts
in the already published AGILE catalogs, 
and do not represent new detections.

\subsection{Spectral analysis}
Tab.~\ref{tab:spectrum_AGLgood} shows the results of the
spectral analysis performed on the most significant sources 
detected in this analysis.
Only the 24 sources detected with a significance $\sqrt{(TS)_{3}}>5$ 
(see Tab.~\ref{tab:good_AGL}, upper part) and $|b|<30^\circ$
have been considered.

For all considered sources, the AGILE spectral index shown in 
the fourth column of Tab.~\ref{tab:spectrum_AGLgood} has 
been calculated generating exposure, counts and diffuse 
background maps over five energy bands: 100$-$200 MeV, 200$-$400 MeV, 
400$-$1000 MeV, 1$-$3 GeV and 3$-$50 GeV, 
under the assumption of a power-law energy distribution.
\par
\begin{figure}[ht]
\begin{center}
\includegraphics[width=1.0\columnwidth]{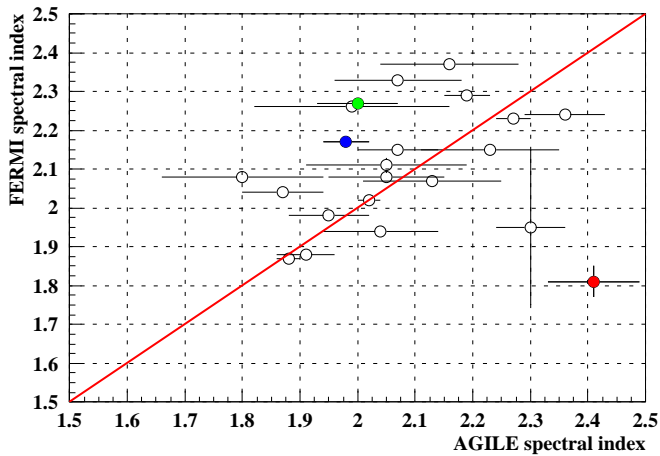}
\end{center}
\caption{Comparison between the spectral index evaluated for the
most significant sources detected by AGILE, and the corresponding
{\it power--law} spectral index reported by Fermi (22 sources).
The horizontal bars represent the AGILE spectral index errors, 
the vertical bars represent the Fermi spectral index errors, when available.
The coloured points refer to the sources highlighted in Fig.~\ref{fig:hpull}
}
\label{fig:spectrum}
\end{figure}
\begin{figure}[ht]
\begin{center}
\includegraphics[width=1.0\columnwidth]{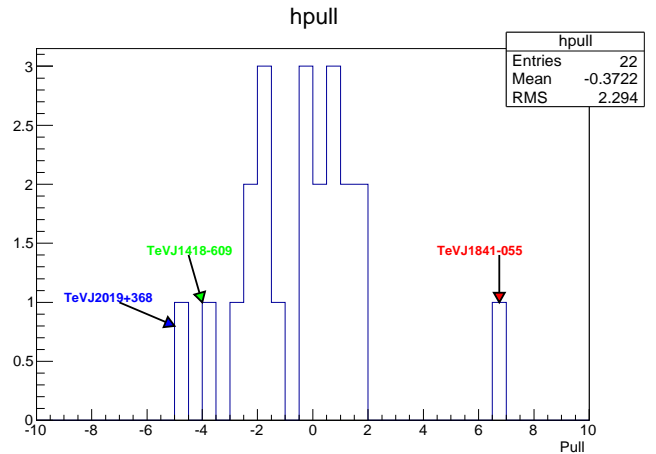}
\end{center}
\caption{Pull distribution of the difference between the AGILE and Fermi 
power-law spectral indices divided by their combined errors. The three sources
with pull larger than three are shown.
}
\label{fig:hpull}
\end{figure}

For comparison, the Fermi power-law spectral index of the 3FGL counterpart 
of the TeV source is also shown in the last column of
the table\footnote{\label{powerlaw} The {\it power law index} reported on 3FGL Catalog \citep{3FGL}
is given without error, and corresponds to the result of fitting the spectrum
with a power-law function; it is equal to {\it Spectral Index} only when spectrum type is
{\tt PowerLaw} and in these cases errors are available.}.
AGILE and Fermi power-law spectral indices are also compared in Fig.~\ref{fig:spectrum}.
A quantitative comparison between Fermi and AGILE power-law spectral indices can be done 
adding quadratically the AGILE and Fermi power-law spectral 
index errors when available or relying only on the AGILE error.
No systematic error is assumed.
{On these very conservative assumptions 
the distribution of the pulls (difference between the
AGILE and Fermi power-law spectral indices divided by their combined errors) 
shown in Fig.\ref{fig:hpull} can be significantly overestimated.
Nevertheless, for most of the sources (19 out of 22), AGILE and Fermi power-law spectral
indices agree within three $\sigma$.}

{As explained in Section 5.2, 
the analysis in this paper was performed taking into account all the known 
$\gamma$-ray sources detected by AGILE at the time of the analysis, and
in general more than one 3FGL source may be within the $95\%$ AGILE error circle 
(plus a suggested systematic error of 0.1$^{\circ}$).
In particular in the TeV J1841-055 case, the one with the largest pull (red),
there are three 3FGL sources within the AGILE detection error circle.
The TeV source has an extension of the order of 0.5 degrees,
and the AGILE spectral result was compared with the associated extended Fermi PWN
3FGL J1840.9-0532e. It is likely that other two $\gamma$-ray sources 
(3FGL J1839.3-0552 and 3FGL J1838.9-0537) contribute to the AGILE detected emission,
both having softer power-law spectral indices.}

\par

\section{Conclusions and discussion}
\label{CONCLUSIONS}
An analysis has been performed on a sample of 152 known TeV sources using the
first two years of AGILE data, with the purpose to detect $\gamma$--ray
emission associated with these TeV sources.

A significant $\gamma$--ray excess in the AGILE data has 
been found for 52 input TeV sources,
corresponding to $\sim$ 34\% of the analysed sample.
In particular 26 new AGILE sources have been found with respect to 
the AGILE reference catalogs, 
15 of which are Galactic, 7 are extragalactic and 4 are 
unidentified. 

Eight of the AGILE detected TeV sources (listed in Sect.~\ref{sec:RESULTS})
have no 3FGL official association and will be further investigated
in a dedicated paper. 

The difference between the two experiments in some cases may be due to 
a different response to softer spectral index sources with spectral energy distribution
peaking in the 100-400 MeV energy range or to differences in the 
assumed background diffuse model.
A new Fermi analysis using the recently delivered ``Pass 8 Data''
providing a significant increase in acceptance at lower energies may result
in new Fermi associations.
Furthermore according to the official
Fermi collaboration association procedure,
only counterparts which reach a posteriori probability of at least $80\%$
are retained, and this does not mean that there is no
3FGL source within the error region. 

The spatial association of a TeV source with an AGILE source 
may be due to chance.
The probability of serendipitous association may be estimated evaluating the 
total sky coverage of all catalogued AGILE sources, defined by their 95\%\ error radius,
and then calculating the overall probability that each TeV source can overlap to
any AGILE source, that is, the distance between a TeV source and an AGILE source is by chance smaller 
than the sum of their error circles \citep{Funk3}.
Because of the density of both TeV and AGILE sources are strongly non uniform, 
being much larger on the Galactic plane, this evaluation is performed separately on 
a narrow stripe along the Galactic plane\footnote{The probability estimation is weakly 
dependent on the boundary used.},
defined by $|b|\le 3^\circ$, and on all the rest of the sky map, defined by $|b| > 3^\circ$.
The number of
serendipitous associations
is found to be 0.82 for such a band around the Galactic
plane, whereas it is 0.08 for the other region.
Therefore the chance coincidence should be $\mathcal{O}(1)$ over all the sky.

It is worth noting that the analysis accomplished and described in this paper
concerns a nearly continuous data taking period of about 2 year, therefore
the obtained fluxes (and corresponding detection significances) are the
average values integrated over a rather long time. For this reason, some
$\gamma$--sources characterised by high variability (for example W Comae \citep{2008ATel.1582}
and 4C +21.35 \citep{2010ATel.2641,2010ATel.2686,2014ATel.6733})
has not be detected in this analysis \citep[and similarly in] [] {AGILEVAR}, even if they have already been
detected and analysed in other previous observations published by AGILE,
referring to shorter time periods in coincidence with their flares.

The majority of the AGILE detected sources are Galactic.
This might be a bias due the higher exposure of the 
Galactic Plane during the Pointing period, see Fig.~\ref{fig:exposure}.

From the ``Analysis flag'' reported in Tab.~\ref{tab:good_AGL},
there are 13 AGILE detected sources in the Galactic plane flagged 
as ``Overlapping'' or ``External'' with the TeV emission region. This may indicate 
a possible displacement between the TeV and the GeV emission regions,
as it is the case for the IC443 SNR~\citep{2010ApJ...710L.151T}. 
%

\section*{Acknowledgments}
\label{ACKNOWLEDGMENTS}
The authors would like to thank the Istituto Nazionale di Astrofisica, 
the Agenzia Spaziale Italiana, the Consorzio Interuniversitario per la 
Fisica Spaziale, and the Istituto Nazionale di Fisica Nucleare 
for their generous support of the AGILE mission and this 
research, including ASI contracts N. I/042/10/1 and I/028/12/0. 


%
%

\bibliographystyle{aa}
\bibliography{TeV_paper}

\onecolumn
\clearpage
\input{TeV_paper_table1_152.tex}

\clearpage

\clearpage
\input{TeV_paper_table2_3fgl.new.tex}
\clearpage

\clearpage
\input{TeV_paper_table3_3fgl.tex}
\clearpage

\clearpage
\input{TeV_paper_table4.new.tex}
\clearpage

\end{document}

%% file: author-list-aa.tex
\author{
A. Rappoldi \inst{1} \thanks{Corresponding author}
\and F. Lucarelli \inst{2,3}
\and C. Pittori \inst{2,3}
\and F. Longo \inst{4,5}
\and P.W. Cattaneo  \inst{1} 
\and F. Verrecchia \inst{2,3}
\and M. Tavani \inst{6,7}
\and A. Bulgarelli \inst{8}
\and A.W. Chen \inst{9}
\and S. Colafrancesco \inst{9}
\and I. Donnarumma \inst{6}
\and A. Giuliani \inst{10}
\and A. Morselli \inst{11}
\and S. Sabatini \inst{6}
\and S. Vercellone \inst{12}
}

\institute{
{INFN-Pavia, via Bassi 6, I-27100 Pavia, Italy} \and
{ASI Science Data Center, via del Politecnico snc 00133, Roma, Italy} \and
{INAF-Osservatorio Astronomico di Roma, via di Frascati 33, I-00040 Monteporzio Catone, Roma, Italy} \and 
{INAF-Osservatorio Astronomico di Trieste, via G.B. Tiepolo 11, I-34143 Trieste, Italy} \and
{INFN-Trieste, Padriciano 99, I-34012 Trieste, Italy} \and
{INAF-IAPS Roma, via Fosso del Cavaliere 100, I-00133 Roma, Italy} \and
{Dip. di Fisica, Universit\`a ``Tor Vergata'', via della Ricerca Scientifica 1, I-00133 Roma, Italy} \and
{INAF-IASF Bologna, via Gobetti 101, I-40129 Bologna, Italy} \and
{School of Physics, University of the Witwatersrand, Johannesburg Wits 2050, South Africa} \and
{INAF-IASF Milano, via E. Bassini 15, I-20133 Milano, Italy} \and
{INFN-Roma Tor Vergata, via della Ricerca Scientifica 1, 00133 Roma, Italy} \and
{INAF-IASF Palermo, Via Ugo La Malfa 153, I-90146  Palermo, Italy} 
}

%% file: TeV_paper_table1_152.tex
\begin{landscape}
{\scriptsize
\LTcapwidth=\textwidth
\begin{longtable}{|c|c|c|c|c|c|c|c|c|c|c|}
\caption{
List of the 152 TeV sources considered in this analysis. The sources passing the detection criteria discussed in Sect.~\ref{sec:detection} are shown in bold. For a detailed description of the table columns please refer to the text.}
\label{tab:all} \\
\hline
ID & {\bf TeV Source} & $(l,b)$ & TeV Pos. Err. & Canonical name & Type \footnotemark 
& $\sqrt{(TS)_4}$ & $\sqrt{(TS)_3}$ & Flux$_4$ (E$>$100~MeV) & UL$_4$ & Ext.\\
& & [deg] \footnotemark & [deg] & & & & & [$\mathrm{10^{-7}\; ph\; cm^{-2}\; s^{-1}}$] & [$\mathrm{10^{-7}\; ph\; cm^{-2}\; s^{-1}}$] & \\
\hline
\endfirsthead
%
\multicolumn{11}{c}%
{{\bfseries \tablename\ \thetable{} -- continued from previous page}} \\
\hline
ID & {\bf TeV Source} & $(l,b)$ & TeV Pos. Err. & Canonical name & Type & $\sqrt{(TS)_4}$ & $\sqrt{(TS)_3}$ & Flux$_4$ (E$>$100~MeV) & UL$_4$ & Ext.\\
& & [deg] \footnotemark & [deg] & & & & & [$\mathrm{10^{-7}\; ph\; cm^{-2}\; s^{-1}}$] & [$\mathrm{10^{-7}\; ph\; cm^{-2}\; s^{-1}}$] & \\
\hline
\endhead
%
\hline \multicolumn{11}{|r|}{{Continued on next page}} \\ \hline
\endfoot
\hline \hline
\endlastfoot

\addtocounter{footnote}{-2} 
\footnotetext{TeV source classification types: PWN: Pulsar Wind Nebulae; BIN: Binary; SNR: SuperNova Remnant; Sbs: StarBursts; UNID: UNIDentified; FSRQ: Flat Spectrum Radio Quasar; HBL: High frequency peaked BL Lac object; IBL: Intermediate frequency peaked BL Lac object; LBL: Low  frequency peaked BL Lac object; XRB: X-Rays Binary; WR: Wolf-Rayet star; FRI: Fanaroff--Riley type I ; GC: Globular Cluster; MC: Molecular Cloud; PSR: Pulsar. Note that the $\gamma$-ray emission detected in the search of counterparts to the TeV 
emission from PWN is in general due to the average (pulsar + nebula) $\gamma$-ray flux values.
Timing analysis of pulsars was not performed in this paper.
}
\addtocounter{footnote}{1} 
\footnotetext{The $^\star$ indicates that the best-fit position of the TeV excess
is not available and the position of the optical/radio known
counterpart has been used.}

{\bf1~~~} & {\bf TeVJ0006+727}&{\bf 119.604, 10.403}&{\bf 0.091}&{\bf CTA 1}&{\bf PWN/SNR} &{\bf 21.4}&{\bf 21.6}&{\bf 3.3$\pm$0.2}& &{\bf X }\\
2  & TeVJ0013-189 & 74.6130, -78.0684  & 0.0083 & SHBL J001355.9-185406	& HBL	&	& &	& 1.5 &  \\
3  & TeVJ0025+640 & 120.106,   1.451   & 0.026  & Tycho SN, G120.1+1.4	& SNR	& 1.6	& 1.6	&  & 0.7 & \\
4  & TeVJ0033-193 & 94.171,   -81.216$^\star$  & -      & KUV 00311-1938	& HBL	& 	& &	& 2.0 &  \\
5  & TeVJ0035+598 & 120.898,  -3.018   & 0.030  & 1ES 0033+595	        & HBL	&	& &	& 0.2 &  \\
6  & TeVJ0047-253 & 97.4696, -87.9672  & 0.0080 & NGC 253	        & Sbs	&	& &	& 0.6 &  \\
7  & TeVJ0112+227 & 129.14,  -39.88$^\star$   & -    & S2 0109+22       & IBL   & 1.8   & 3.3   &  & 0.8 & \\
8  & TeVJ0136+391 & 132.416, -22.940$^\star$  & -    & RGB J0136+391	& HBL	& 1.2	& 1.2	&  & 0.5 & \\%
9  & TeVJ0152+017 & 152.343, -57.561   & 0.030  & RGB J0152+017	        & HBL	& 1.9	& 3.7	&  & 1.5 & \\%
10 & TeVJ0209+648 & 130.701,  3.102    & 0.059  & 3C 58                 & PWN   & 2.1   & 4.8   &  & 0.8 & \\
11 & TeVJ0218+359 & 142.602, -23.487   & -      & S3 0218+35            & AGN   & 3.3   & 3.9   &  & 1.0 & \\
{\bf 12}&{\bf TeVJ0222+430}\footnote{\label{id9}The sources TeVJ0222+430 and TeVJ0223+430 are as close as 0.1$^\circ$ and therefore undistinguishable in this analysis. Their detection is counted as one.}&{\bf 140.143, -16.767}&{\bf -}&{\bf 3C 66A}&{\bf IBL}&{\bf 8.0}&{\bf 8.1}&{\bf 1.4$\pm$0.2}& & \\
13 & TeVJ0223+430\footref{id9} & 140.254, -16.772 & 0.048 & MAGIC J0223+430& UNID& 7.6 & 8.0 & 1.3$\pm$0.2 & &\\
{\bf 14}&{\bf TeVJ0232+202}&{\bf 152.970, -36.613}&{\bf 0.014}&{\bf 1ES 0229+200}&{\bf HBL}&{\bf 3.1}&{\bf 4.2}&{\bf 0.9$\pm$0.3} & &\\
{\bf 15}&{\bf TeVJ0240+612}&{\bf 135.668, 1.113}&{\bf 0.034}&{\bf LSI+61\_303}&{\bf XRB}&{\bf 26.7}&{\bf 27.0}&{\bf 6.5$\pm$0.3}& &\\
16 & TeVJ0303-241 & 214.6296, -60.1899  & 0.0073 & PKS 0301-243	        & HBL	& 	& &	& 0.9 &  \\
17 & TeVJ0316+413 & 150.183,  -13.734   & -      & IC 310	        & HBL	&       & &	& 0.3 &  \\
18 & TeVJ0319+187 & 165.088,  -31.708$^\star$  & 0.034  & RBS 0413	        & HBL	& 2.0	& &	& 1.2 & \\
{\bf 19}&{\bf TeVJ0319+415}&{\bf 150.576, -13.261}&{\bf -}&{\bf NGC 1275}&{\bf FRI}&{\bf 5.5 }&{\bf 5.5}&{\bf 1.0$\pm$0.2}& &\\
20 & TeVJ0349-119 & 201.909,  -45.704   & 0.011  & 1ES 0347-121	        & HBL	& &	& & 0.8 &  \\
21 & TeVJ0416+010 & 191.8167, -33.1581  & 0.0069 & 1ES 0414+009	        & HBL	& 2.4	& 3.4	&  & 1.5 & \\
22 & TeVJ0449-438 & 248.8066, -39.9082  & 0.0072 & PKS 0447-439	        & HBL	& 1.3	& 1.3	&  & 1.1 & \\
23 & TeVJ0507+676 &  143.795,   15.890$^\star$  & -      & 1ES 0502+675	        & HBL	&	& &     & 0.2 & \\
{\bf 24}&{\bf TeVJ0521+211}\footnote{\label{id21}Results of a refined MLE analysis performed using the updated AGILE best-fit position of the Crab Nebula (one of the known $\gamma$--ray sources within 10 deg from the TeV J0521+211 position) obtained in this work.} & {\bf 183.6021, -8.7114}&{\bf 0.0080}&{\bf RGB J0521.8+211}&{\bf IBL}&{\bf 4.5}&{\bf 4.7}&{\bf 1.6$\pm$0.4} & &\\
25 & TeVJ0525-696 & 280.307,   -32.784$^\star$   & -      & LMC N132D           & SNR/MC& 1.0   & 1.0   &  & 0.6 & \\
{\bf 26}&{\bf TeVJ0534+220}&{\bf 184.5558, -5.7870}&{\bf 0.0070}&{\bf Crab Nebula}&{\bf PWN}&{\bf 55.7}&{\bf 55.9}&{\bf 26.7$\pm$0.7}& &\\
27 & TeVJ0535-692 & 279.60,    -31.91    & 0.022  & 30 Dor C      & Superbubble & 3.5   & 3.8   &  & 1.1 & \\
28 & TeVJ0537-691 & 279.553,   -31.750   & 0.010  & N 157B	        & PWN	& 3.4	& 3.7	&  & 1.1 & \\
29 & TeVJ0550-322 & 237.562,   -26.152   & 0.014  & PKS 0548-322        & HBL	& 	& 	&  & 0.6 & \\
{\bf 30}&{\bf TeVJ0616+225}&{\bf 189.073, 2.918}&{\bf 0.085}&{\bf IC 443}&{\bf SNR}&{\bf 12.9}&{\bf 13.2}&{\bf 5.0$\pm$0.5}& &{\bf X}\\
31 & TeVJ0632+057 & 205.660,    -1.441   & 0.010  & HESS J0632+057      & XRB   & 2.5	& 4.8 	&  & 1.9 & \\
{\bf 32}&{\bf TeVJ0632+173}& {\bf 195.34, 3.78}&{\bf 0.50}&{\bf Geminga PWN}&{\bf PWN}&{\bf 75.2}&{\bf 82.6}&{\bf 40.3$\pm$0.9}& &{\bf X}\\
33 & TeVJ0648+152 &  198.99,      6.32   & 0.12  & RX J0648.7+1516	& HBL	& 	& 	&  & 0.7 & \\
34 & TeVJ0650+250 & 190.282,     10.996$^\star$  & -     & 1ES 0647+250	        & HBL	& 	&	&  & 0.6 & \\
35 & TeVJ0710+591 & 157.391,     25.421  & 0.027 & RGB J0710+591       	& HBL	& 	&	&  & 0.4 & \\
{\bf 36}&{\bf TeVJ0721+713}&{\bf 143.981, 28.018}$^\star$&{\bf -}&{\bf S5 0716+714}&{\bf LBL}&{\bf 13.8}&{\bf 13.9}&{\bf 2.6$\pm$0.2}& &\\
37 & TeVJ0809+523 & 166.246,     32.935  & 0.048 & 1ES 0806+524         & HBL	& 1.9	& 2.0	&  & 0.9 & \\
{\bf 38} & {\bf TeVJ0835-455}\footnote{\label{VelaX} The TeV error region for the Vela X PWN does not overlap the AGILE Vela Pulsar position error, hence the Vela PSR average $\gamma$--ray flux value for a spectral index of -1.69 was subtracted in the MLE automatic analysis.
However the result is affected by the very intense nearby Vela Pulsar. The Vela X accurate $\gamma$--ray flux estimate and 
source location by AGILE is discused in the dedicated analysis in \citep{Pellizzoni:2009ey}.} & {\bf 263.840, -3.073} & {\bf 0.034} & {\bf Vela X} & {\bf PWN} & {\bf7.9} & {\bf 11.2} & {\bf 6.2$\pm$0.8}& & {\bf X}\\
39 & TeVJ0847+115 & 215.456,     30.890$^\star$  & - 	 & RBS 0723	        & HBL	&       & 	&  & 0.8 & \\
{\bf 40} & {\bf TeVJ0852-463}\footnote{\label{VelaJr}Also in this case, the MLE automatic analysis is 
affected by the very intense nearby Vela Pulsar. A dedicated analysis will be performed.} & {\bf 266.285, -1.241} & {\bf -} & {\bf RX J0852.0-4622} & {\bf SNR}& {\bf 4.0} & {\bf 8.5} & {\bf 1.5$\pm$0.4} & & {\bf X} \\
41 & TeVJ0955+696 & 141.409,     40.568$^\star$  & - 	 & M82                  & Sbs	&	&	&  & 0.3 & \\
42 & TeVJ0958+655 & 145.75,      43.13$^\star$   & -     & S4 0954+65           & FSRQ  & 1.8   & 1.9   &  & 0.8 & \\
43 & TeVJ1010-313 & 266.896,     20.063  & 0.017 & 1RXS J101015.9-311909& HBL	& 	& 	&  & 0.1 & \\
44 & TeVJ1015+494 & 165.534,     52.712$^\star$  & - 	 & 1ES 1011+496	        & HBL	& 2.4	& 2.9	&  & 1.2 & \\
{\bf 45}&{\bf TeVJ1018-589}&{\bf 284.256, -1.818}&{\bf -}&{\bf HESS J1018-589}&{\bf BIN}&{\bf 7.0}&{\bf 7.9}&{\bf 2.4$\pm$0.4}& &{\bf X}\\
46 & TeVJ1023-575 & 284.217,     -0.401  & 0.030 & Westerlund 2	        & WR	& 2.1	& 3.5	&  & 1.5 & X \\
47 & TeVJ1026-582 & 284.798,     -0.520  & 0.090 & HESS J1026-582       & PWN	&	&	&  & 0.2 & X\\
48 & TeVJ1103-234 & 273.188,     33.074  & 0.012 & 1ES 1101-232	        & HBL	& 	&	&  & 0.3 & \\
{\bf 49}&{\bf TeVJ1104+382}&{\bf 179.832, 65.032}&{\bf -}&{\bf Mrk 421}&{\bf HBL}&{\bf 5.2}&{\bf 5.2}&{\bf 1.6$\pm$0.4}& & \\
50 & TeVJ1119-614 & 292.102,      -0.487 & - 	 & G 292.2-0.5	        & PWN  & 2.3	& 4.6	&  & 1.3 & X\\
51 & TeVJ1136+676 & 133.453,      47.951$^\star$ & -     & RXJ1136.5+6737       & HBL  &       &       &  & 0.4 & \\
52 & TeVJ1136+701 & 131.910,      45.641$^\star$ & -     & Mrk 180 	        & HBL	& 1.4	& 1.4	&  & 0.7 &\\
53 & TeVJ1217+301 & 189.010,      82.046 & 0.016 & 1ES 1215+303         & HBL	& 	&	&  & 0.3 & \\
54 & TeVJ1221+282\footnote{Detected by AGILE during a flaring episode~\citep{2008ATel.1582}.} & 201.734, 83.288  & - & W Comae & IBL	&  & & & 1.0 &  \\
55 & TeVJ1221+301 & 186.210,      82.743 & 0.031 & 1ES 1218+304         & HBL	& 	&	&  & 0.2 & \\
56 & TeVJ1224+213\footnote{Detected by AGILE during several flaring episodes~\citep{2010ATel.2641,2010ATel.2686,2014ATel.6733}.} & 255.074, 81.660$^\star$ & -	   & 4C +21.35 & FSRQ & 2.6 & 3.5 & & 1.7 & \\
57 & TeVJ1224+246 & 233.952,      83.418 & -	 & MS 1221.8+2452	& HBL	&	&	&  & 0.7 & \\
58 & TeVJ1230+123 & 283.7388,    74.4946$^\star$ & 0.0080& M 87         & FRI	& 	&	&  & 1.0 & \\
59 & TeVJ1230+253 & 232.75,      84.91$^\star$   &  -    & S3 1227+25   & IBL   & 2.8   & 3.0   &  & 1.7 & \\
{\bf 60}&{\bf TeVJ1256-057}&{\bf 305.104, 57.062}$^\star$&{\bf -}&{\bf 3C 279}&{\bf FSRQ}&{\bf 11.1}&{\bf 11.2}&{\bf 4.2$\pm$0.5}& & \\
61 & TeVJ1302-638 & 304.187,      -0.987 & 0.011 & PSR B1259-63	        & XRB	&  	& 	&  & 0.6 & \\
62 & TeVJ1303-631 & 304.213,      -0.334 & 0.014 & HESS J1303-631	& PWN	& 	&	&  & 0.4 & X\\
63 & TeVJ1315-426 & 307.540,      20.064 & 0.018 & 1ES 1312-423	        & HBL	& 2.4	& 5.1	&  & 0.9 &\\
{\bf 64}&{\bf TeVJ1325-430}\footnote{\label{CenA}The region of the MLE analysis is not yet well modelled since this source shows a point-like core and extended emission from the lobes, which are not yet included in the AGILE reference Catalogs. A dedicated analysis will be performed.} & {\bf 309.513, 19.425}&{\bf 0.021}&{\bf Centaurus A}&{\bf FRI}&{\bf 4.4}&{\bf 4.9}&{\bf 0.8$\pm$0.2}& & \\
65 & TeVJ1356-645 & 309.812,      -2.494 & 0.034 & HESS J1356-645	& PWN	& 1.7	& 2.0	&  & 1.0 & X\\%
{\bf 66}&{\bf TeVJ1418-609}&{\bf 313.247, 0.150}&{\bf 0.030}&{\bf Kookaburra (Rabbit)}&{\bf PWN}&{\bf 13.5}&{\bf 13.5}&{\bf 5.1$\pm$0.4}& &{\bf X}\\
67 & TeVJ1420-607 & 313.558,       0.268 & 0.018 & Kookaburra (PWN)	& PWN	& 	& & & 1.1 & X\\
68 & TeVJ1427+238 &  29.472,      68.208 & 0.033 & PKS 1424+240	        & IBL	& 1.3	& 1.3 & & 1.3 & \\
69 & TeVJ1427-608 & 314.408,      -0.145 & 0.050 & HESS J1427-608	& UNID	& 1.2	& 1.2 & & 1.1 & \\
70 & TeVJ1428+426 &  77.487,      64.899$^\star$ & -	 & H 1426+428	        & HBL	& 	& & & 1.0 & X \\
71 & TeVJ1442-624 & 315.410,      -2.300 & 0.059 & RCW 86               & SNR	& 	& & & 0.5 & X \\
72 & TeVJ1443+120 &   8.330,      59.840$^\star$ & - 	 & 1ES 1440+122		& IBL	& 	& & & 0.6 & \\
73 & TeVJ1443+250 &  34.56,       64.70$^\star$  &  -    & PKS 1441+25          & FSRQ  &       & & & 0.7 & \\
74 & TeVJ1457-594 &  318.36,       -0.43 & 0.14  & G 318.2+0.1	       	& SNR/MC&  	& & & 0.2 & X \\
{\bf 75}&{\bf TeVJ1459-608}&{\bf 317.748, -1.704}&{\bf 0.030}&{\bf HESS J1458-608}&{\bf PWN}&{\bf 5.6}&{\bf 5.8}&{\bf 1.5$\pm$0.3}& &{\bf X}\\ 
76 & TeVJ1502-419 & 327.580,      14.571 & -	 & SN 1006	        & SNR	& 	& & & 0.3 & \\
77 & TeVJ1503-582 &  319.62,        0.29 & 0.10  & HESS J1503-582	& UNID	& 	& & & 0.2 & X \\
78 & TeVJ1506-623 & 317.946,      -3.494 & 0.050 & HESS J1507-622	& UNID	& 3.1	& 4.3 & & 1.2 & X\\
{\bf 79}&{\bf TeVJ1512-091}&{\bf 351.2907, 40.1296}&{\bf 0.0093}&{\bf PKS 1510-089}&{\bf FSRQ}&{\bf 24.8}&{\bf 25.0}&{\bf 8.1$\pm$0.4}& & \\
80 & TeVJ1514-591 & 320.324,      -1.200 & 0.010 & MSH 15-52          	& PWN	& 	& & & 0.6 & X \\
81 & TeVJ1517-243 & 340.673,      27.577$^\star$ & 0.014 & AP Lib             	& LBL	& 2.0	& 3.6 & & 0.8 & \\
82 & TeVJ1554-550 & 327.158,      -1.072 & 0.018 & G 327.1-1.1 		& PWN	& 	& & & 0.3 & X \\
83 & TeVJ1555+111 &  21.919,      43.960 & 0.016 & PG 1553+113		& HBL	& 1.9	& 4.4 & & 1.3 & \\
84 & TeVJ1614-518 &  331.52,       -0.58 & -     & HESS J1614-518	& UNID	&    	& & & 0.8 & X \\
85 & TeVJ1616-508 &  332.39,       -0.14 & -     & HESS J1616-508	& PWN	& 	& & & 0.9 & X \\
86 & TeVJ1626-490 & 334.772,       0.045 & 0.050 & HESS J1626-490	& UNID	& 	& & & 0.3 & \\
{\bf 87}&{\bf TeVJ1632-478}\footnote{\label{icrc11}AGILE detection presented at the 32nd ICRC~\citep{2011ICRC}.}&{\bf 336.38, 0.19}&{\bf -}&{\bf HESS J1632-478}&{\bf UNID}&{\bf 5.6}&{\bf 5.7}&{\bf 2.2$\pm$0.4}& &{\bf X}\\
{\bf 88}&{\bf TeVJ1634-472}&{\bf 337.11, 0.22}&{\bf -}&{\bf HESS J1634-472}&{\bf UNID}&{\bf 15.4}&{\bf 15.7}&{\bf 6.3$\pm$0.4}& &{\bf X}\\
{\bf 89}&{\bf TeVJ1640-465}&{\bf 338.32, -0.02}&{\bf -}&{\bf HESS J1640-465}&{\bf PWN}&{\bf 12.3}&{\bf 15.2}&{\bf 5.1$\pm$0.5}& &{\bf X}\\
90 & TeVJ1641-463 & 338.519,       0.095 & 0.015 & HESS J1641-463	& UNID	& 	& & & 1.0 & \\
91 & TeVJ1647-458 &  339.55,       -0.35 & 0.12  & Westerlund 1		& WR    & 1.3	& 1.3 & & 1.3 & X \\
92 & TeVJ1653+397 &  63.600,      38.859 & - 	 & Mrk 501		& HBL   & 3.9	& 5.3 & & 1.3 & \\
93 & TeVJ1702-420 & 344.304,      -0.184 & 0.050 & HESS J1702-420	& UNID	& 	& & & 0.2 & X \\
94 & TeVJ1708-410 & 345.683,      -0.469 & 0.050 & HESS J1708-410	& UNID	& 	& & & 0.4 & X \\
{\bf 95}&{\bf TeVJ1708-443}&{\bf 343.058, -2.376}&{\bf 0.071}&{\bf HESS J1708-443}&{\bf PWN/SNR}&{\bf 39.8}&{\bf 42.3}&{\bf 13.6$\pm$0.4}& &{\bf X}\\
{\bf 96}&{\bf TeVJ1713-382}&{\bf 348.639, 0.388}&{\bf 0.018}&{\bf CTB 37B}&{\bf SNR}&{\bf 6.5}&{\bf 7.8}&{\bf 2.6$\pm$0.4} & &{\bf X}\\
97 & TeVJ1713-397 & 347.336,      -0.473 & -     & RX J1713.7-3946 	& SNR   & 3.7	& 7.0 & & 2.2 & X \\
{\bf 98}&{\bf TeVJ1714-385}\footnote{\label{MLconfused}Source located in a crowded region of the Galactic Plane. The automatic MLE analysis is not reliable. A dedicated analysis will be performed.}&{\bf 348.389, 0.107}&{\bf 0.023}&{\bf CTB 37A}&{\bf SNR}&{\bf 7.0}&{\bf 7.5}&{\bf 2.8$\pm$0.4}& &{\bf X}\\%
{\bf 99}&{\bf TeVJ1718-374}&{\bf 349.720, 0.174}&{\bf 0.010}&{\bf G 349.7+0.2}&{\bf SNR/MC}&{\bf 6.2}& {\bf 7.6}& {\bf 2.6$\pm$0.4}& & \\
{\bf 100}&{\bf TeVJ1718-385}&{\bf 348.834, -0.488}&{\bf 0.034}&{\bf HESS J1718-385}&{\bf PWN}&{\bf 6.9}&{\bf 7.5}&{\bf 2.7$\pm$0.4}& &{\bf X}\\
101 & TeVJ1725+118 &  34.120,      24.475$^\star$ & - 	 & H 1722+119	        & HBL	& 	& & & 0.4 & \\
102 & TeVJ1728+502 &  77.068,      33.537 & - 	 & 1ES 1727+502	        & HBL	&	& & & 0.4 & \\
{\bf 103}&{\bf TeVJ1729-345}&{\bf 353.444, -0.128}&{\bf 0.035}&{\bf HESS J1729-345}&{\bf UNID}&{\bf 8.0}&{\bf 10.3}&{\bf 3.6$\pm$0.5}& &{\bf X}\\
{\bf 104}&{\bf TeVJ1732-347}\footref{icrc11} & {\bf 353.542, -0.670}&{\bf -}&{\bf HESS J1731-347}&{\bf SNR}&{\bf 7.3}&{\bf 10.2}&{\bf 3.1$\pm$0.5}& &{\bf X}\\
{\bf 105}&{\bf TeVJ1741-302}\footnote{\label{gc}Source located in the region near the Galactic Center. The automatic MLE analysis is not reliable. A dedicated analysis is being performed with an improved AGILE diffuse background model in the Galactic Center region (Fioretti et al., in preparation).}&{\bf 358.397, 0.191}&{\bf -}&{\bf HESS J1741-302}&{\bf UNID}&{\bf 4.3}&{\bf 9.6}&{\bf 2.1$\pm$0.5}& &\\
106 & TeVJ1743+196 & 43.836,      23.339$^\star$ & - 	 & 1ES 1741+196	        & HBL	& 	& & & 0.3 & \\
107 & TeVJ1745-290 & 359.9449,   -0.0440 & 0.0024& HESS J1745-290	& UNID	&	& & & 0.3 & \\
{\bf 108}&{\bf TeVJ1745-303}&{\bf 358.710, -0.640}&{\bf -}&{\bf HESS J1745-303}&{\bf SNR/MC}&{\bf 11.2}&{\bf 12.7}&{\bf 5.5$\pm$0.5}& & {\bf X}\\
{\bf 109}&{\bf TeVJ1747-248}&{\bf 3.78, 1.72}&{\bf 0.45}&{\bf Terzan 5}&{\bf GC}&{\bf 5.2}&{\bf 5.9}&{\bf 1.9$\pm$0.4}& & {\bf X}\\
110 & TeVJ1747-281 &   0.872,       0.076 & -     & G 0.9+0.1	        & PWN	& 	& & & 0.1 & \\
111 & TeVJ1800-240 &   5.960,      -0.380 & 0.044 & HESS J1800-240 (A+B+C) & UNID & 	& & & 0.9 & X \\
{\bf 112}&{\bf TeVJ1801-233}&{\bf 6.657, -0.268}&{\bf 0.032}&{\bf W28}&{\bf SNR/MC}&{\bf 14.0}&{\bf 15.0}&{\bf 6.5$\pm$0.5}& & {\bf X}\\
{\bf 113}&{\bf TeVJ1804-216}&{\bf 8.354, -0.000}&{\bf -}&{\bf HESS J1804-216}&{\bf UNID}&{\bf 7.5}&{\bf 7.8}&{\bf	3.4$\pm$0.5}& &{\bf X}\\
114 & TeVJ1808-204 &  9.960,      -0.248 & 0.038 & HESS J1808-204       & UNID	&	& & & 1.3 & \\
115 & TeVJ1809-193 & 11.180,      -0.088 & 0.050 & HESS J1809-193	& PWN	& 3.4	& 4.1 & & 2.4 & X\\
{\bf 116}&{\bf TeVJ1813-178}& {\bf 12.812, -0.026}&{\bf -}&{\bf HESS J1813-178}&{\bf PWN}&{\bf 4.1}&{\bf 4.8}&{\bf 1.9$\pm$0.5}& & {\bf X}\\
117 & TeVJ1818-154 & 15.409,       0.161 & 0.014 & G 15.4+0.1	        & PWN	& 1.8	& 9.0 & & 1.7 & X \\
{\bf 118}&{\bf TeVJ1825-137}&{\bf 17.711, -0.697}&{\bf 0.018}&{\bf HESS J1825-137}&{\bf PWN}&{\bf 8.0}&{\bf 11.3}&{\bf 3.8$\pm$0.5}& & {\bf X}\\
{\bf 119}&{\bf TeVJ1826-148}\footref{MLconfused} & {\bf 16.902, -1.278} & {\bf 0.012}&{\bf LS 5039}&{\bf XRB}&{\bf 7.7} & {\bf 10.7} & {\bf 3.3$\pm$0.5} & & \\
120 & TeVJ1831-099 & 21.850,      -0.109 & -	 & HESS J1831-098	& PWN	& 1.9	& 4.2 & & 1.8 & X \\
121 & TeVJ1832-093 & 22.476,      -0.177 & 0.015 & G 22.7-0.2	        & SNR/MC& 1.1	& 1.1 & & 1.4 & X \\
122 & TeVJ1833-106 & 21.511,      -0.876 & 0.016 & HESS J1833-105	& UNID	& 2.3	& 3.6 & & 1.8 & \\
123 & TeVJ1834-087 &  23.24,       -0.32 & - 	 & HESS J1834-087  	& UNID	& 	& & & 0.9 & X \\
{\bf 124}&{\bf TeVJ1837-069}&{\bf 25.18, -0.11}&{\bf -}&{\bf HESS J1837-069}&{\bf UNID}&{\bf 4.9}&{\bf 7.4}&{\bf 2.1$\pm$0.4}& &{\bf X}\\
{\bf 125}&{\bf TeVJ1841-055}\footref{icrc11}&{\bf 26.795, -0.198}&{\bf 0.050}&{\bf HESS J1841-055}&{\bf UNID}&{\bf 12.4}&  {\bf 14.0}&{\bf 5.0$\pm$0.4}& &{\bf X}\\
126 & TeVJ1843-030 & 29.033,       0.370 & -	 & HESS J1843-033	& UNID	& 3.6	& 6.7 & & 2.2 & \\
127 & TeVJ1846-029 & 29.705,      -0.240 & 0.011 & HESS J1846-029       & PWN	& 1.6	& 1.6 & & 1.5 & \\
{\bf 128}&{\bf TeVJ1848-017}&{\bf 31.000, -0.160}&{\bf -}&{\bf WR121a/W43}&{\bf WR}&{\bf 3.7}&{\bf 4.6}&{\bf 1.4$\pm$0.4}& &{\bf X}\\
129 & TeVJ1849-000 & 32.638,       0.526 & -	 & IGR J18490-0000	& PWN	& 	& & & 1.0 & \\
130 & TeVJ1857+026 & 36.003,      -0.061 & 0.039 & HESS J1857+026	& UNID	& 	& & & 0.2 & X \\
131 & TeVJ1858+020 & 35.578,      -0.581 & 0.050 & HESS J1858+020	& UNID	&	& & & 0.1 & X \\
{\bf 132}&{\bf TeVJ1907+062}&{\bf 40.280, -0.688}&{\bf 0.020}&{\bf MGRO J1908+06}&{\bf UNID}&{\bf 12.9}&{\bf 13.3}&{\bf 4.5$\pm$0.4}& &{\bf X}\\
{\bf 133}&{\bf TeVJ1911+090}&{\bf 43.259, -0.189}&{\bf 0.071}&{\bf W 49B}&{\bf SNR/MC}&{\bf 7.6}&{\bf 7.8}&{\bf 2.4$\pm$0.3}& & \\
{\bf 134}&{\bf TeVJ1912+101}&{\bf 44.391, -0.071}&{\bf 0.050}&{\bf HESS J1912+101}&{\bf PWN}&{\bf 4.1}&{\bf 7.6}&{\bf 1.3$\pm$0.3}& &{\bf X}\\
{\bf 135}&{\bf TeVJ1923+141}&{\bf 49.116, -0.365}&{\bf 0.015}&{\bf W 51}&{\bf SNR/MC}&{\bf 7.0}&{\bf 7.1}&{\bf 2.1$\pm$0.3}& &{\bf X}\\
136 & TeVJ1930+188 &  54.10,        0.26 & 0.11  & G 54.1+0.3           & PWN	& 3.5	& 5.2 & & 1.5 & \\
137 & TeVJ1943+213 &57.7577,     -1.2928 & 0.0073& HESS J1943+213	& HBL	& 	& & & 0.5 & \\
138 & TeVJ1959+651 & 98.003,      17.670$^\star$ & - 	 & 1ES 1959+650         & HBL   & 3.4	& 3.7 & & 0.7 & \\
{\bf 139}&{\bf TeVJ2001+438}&{\bf 79.071, 7.110}&{\bf -}&{\bf MAGIC J2001+435}&{\bf HBL}&{\bf 4.9}&{\bf 4.8}&{\bf 0.9$\pm$0.2} & & \\
140 & TeVJ2009-488 &350.3741,   -32.6052 & 0.0083& PKS 2005-489         & HBL	& 	& & & 0.4 & \\
141 & TeVJ2016+372 &  74.940,      1.140 & 0.019 & VER J2016+372	& UNID	&	& & & 0.3 & \\
{\bf 142}&{\bf TeVJ2019+368}&{\bf 74.828, 0.417}&{\bf 0.090}&{\bf MGRO J2019+37}&{\bf PWN}&{\bf 24.7}&{\bf 25.5}&{\bf 6.7$\pm$0.3}& & {\bf X}\\
143 & TeVJ2019+407 &  78.331,      2.489 & 0.035 & VER J2019+407	& UNID	& 	& & & 0.1 & X \\
{\bf 144}&{\bf TeVJ2032+415}&{\bf 80.279, 1.042}&{\bf 0.044}&{\bf TeV J2032+4130}&{\bf UNID}&{\bf 7.6}&{\bf 8.0}&{\bf 2.2$\pm$0.3}& &{\bf X}\\
{\bf 145}&{\bf TeVJ2158-302}&{\bf 17.737, -52.247}&{\bf -}&{\bf PKS 2155-304}&{\bf HBL}&{\bf 4.6}&{\bf 4.8}&{\bf 1.3$\pm$0.3}& & \\
{\bf 146}&{\bf TeVJ2202+422}&{\bf 92.590, -10.441}&{\bf -}&{\bf BL Lacertae}&{\bf	LBL}&{\bf 7.2}&{\bf 8.0}&{\bf 1.0$\pm$0.2}& & \\
{\bf 147}&{\bf TeVJ2227+608}&{\bf 106.35, 2.71}&{\bf 0.10}&{\bf G 106.3+2.7}&{\bf SNR}&{\bf 15.2}&{\bf 16.7}&{\bf 3.0$\pm$0.2}& &{\bf X}\\
148 & TeVJ2243+203 & 86.567,     -33.365$^\star$ & -    & RGB J2243+203         & HBL   &       & & & 0.4 & \\
149 & TeVJ2250+384 & 98.254,     -18.578 & -	& B3 2247+381           & HBL	& 	& & & 0.2 & \\
{\bf 150}&{\bf TeVJ2323+588}&{\bf 111.735, -2.130}&{\bf 0.020}&{\bf Cassiopeia A}&{\bf SNR}&{\bf 5.2}&{\bf 7.2}&{\bf 0.9$\pm$0.2}& & \\
151 & TeVJ2347+517 & 112.73,       -9.86 & 0.10 & 1ES 2344+514	        & HBL	 & 	   &         &                 & 0.3 & \\
{\bf 152}&{\bf TeVJ2359-306}&{\bf 12.8689, -78.0367}&{\bf 0.0086}&{\bf H 2356-309}   &{\bf HBL}&{\bf 3.9}&{\bf 4.0}&{\bf 3.3$\pm$1.0}& & \\
\hline
\end{longtable}
}
\end{landscape}

%% file: TeV_paper_table2_3fgl.new.tex
\begin{landscape}
{\scriptsize
\LTcapwidth=1.1\textwidth
\begin{longtable}{|c|c|c|c|c|c|c|c|c|c|}
\caption{
Results for all the sources detected with AGILE in this work, according to the criteria 
described in the text. The upper part reports the results of the MLE analysis 
{\em step 3}, and the lower part the results of the {\em step 4}.}
\label{tab:good_AGL} \\
\hline
ID  &  {\bf TeV Source}  &  $\sqrt{(TS)}$  &  $(l,b)$  &  Error (95\%)\footnote{The AGILE Team recommends to add linearly a systematic error of $\pm 0.1^\circ$.}  &  Flux (E$>$100~MeV)  &  Dist.  &  AGILE  &  Fermi  &  Analysis \\
 &   &   &  [deg]  &  [deg]  &  [$\mathrm{10^{-7}\; ph\; cm^{-2}\; s^{-1}}$]  &  [deg]  &  Association  &  Association  &  Flag\\
\hline 
1	 & 	TeVJ0006+727	 & 	21.6	 & 	119.66,	10.51	 & 0.09 & 	3.3	$\pm$	0.2	 & 0.1 &  1AGLR J0007+7307  &  3FGL J0007.0+7302 (E)  &  IN \\
12     & 	TeVJ0222+430	 & 	8.1	 & 	140.0,	-16.7	 & 0.2 & 	1.4	$\pm$	0.2	 & 0.1 & 	1AGLR J0222+4305	 &  3FGL J0222.6+4301 (P)	 &  IN \\
14     &      TeVJ0232+202       &      4.2    &      152.9,     -36.3   & 0.6 &      1.1   $\pm$ 0.3    & 0.4 &  -  &  3FGL J0232.8+2016 (P)  &  IN \\
15     &      TeVJ0240+612       &      27.1   &      135.5,     1.2     & 0.1 &      6.6   $\pm$ 0.3    & 0.2 &      1AGLR J0240+6115   &  3FGL J0240.5+6113 (P)    &  E \\
19     &      TeVJ0319+415       &      5.5    &      150.6,     -13.2   & 0.4 &      1.0   $\pm$ 0.2    & 0.1 &      1AGLR J0321+4137   &  3FGL J0319.8+4130 (P)    &  IN \\
24     &      TeVJ0521+211       &      4.7    &      183.6,     -8.6    & 0.5 &      1.7   $\pm$ 0.4    & 0.1 &      -      &  3FGL J0521.7+2113 (P)    &  IN \\
26     &      TeVJ0534+220       &      55.9   &      184.48,    -5.81   & 0.06&      26.7  $\pm$ 0.7    & 0.1 &      1AGL J0535+2205  &  3FGL J0534.5+2201 (P)    &  IN \\
30     &      TeVJ0616+225       &      13.2   &      188.9,      3.0    & 0.2 &      5.0   $\pm$ 0.5    & 0.2 &      1AGL J0617+2236    &  3FGL J0617.2+2234e (E)   &      O \\
32     &      TeVJ0632+173       &      82.6   &      195.09,     4.28   & 0.04&      41.8  $\pm$ 0.9    & 0.6 &      1AGL J0634+1748    &  3FGL J0633.9+1746 (E)    &  IN \\
36     &      TeVJ0721+713       &      13.9   &       143.9,     28.1   & 0.1 &      2.6   $\pm$ 0.2    & 0.1 &      1AGLR J0723+7121   &  3FGL J0721.9+7120 (P)    &  IN \\
45     &      TeVJ1018-589       &      7.8    &       284.0,     -2.0   & 0.3 &      2.6   $\pm$ 0.4    & 0.3 &      1AGLR J1018-5852   &  $ \begin{cases} \textrm{3FGL J1018.9-5856 (E)} \\ \textrm{3FGL J1016.3-5858 (E)} \end{cases} $  &  O \\
49     &      TeVJ1104+382       &      5.2    &       179.7,     65.0   & 0.2 &      1.6   $\pm$ 0.4    & 0.1 &      1AGLR J1105+3818   &  3FGL J1104.4+3812 (P)    &  IN \\
60     &      TeVJ1256-057       &      11.2   &       305.3,     57.1   & 0.2 &      4.2   $\pm$ 0.5    & 0.1 &      1AGL J1256-0549    &  3FGL J1256.1-0547 (P)    &  IN \\
66     &      TeVJ1418-609       &      13.5   &       313.2,     0.1    & 0.1 &      5.1   $\pm$ 0.4    & 0.1 &      1AGLR J1417-6108   &  3FGL J1418.6-6058 (E)  &  IN \\
75     &      TeVJ1459-608       &      5.8    &       317.6,     -1.7   & 0.3 &      1.6   $\pm$ 0.3    & 0.1 &      -      &  $ \begin{cases} \textrm{3FGLJ1456.7-6046 (E)} \\ \textrm{3FGLJ1459.4-6053 (E)} \end{cases} $  &  IN \\
79     &      TeVJ1512-091       &      25.0   &       351.4,      40.1  & 0.1 &      8.2   $\pm$ 0.4    & 0.1 &      1AGLR J1513-0906   &  3FGL J1512.8-0906 (P)    &  IN \\
87     &      TeVJ1632-478       &      5.7    &       336.4,      0.0   & 0.4 &      2.2   $\pm$ 0.4    & 0.2 &      -        &  3FGL J1633.0-4746e (E)     &  IN \\
88     &      TeVJ1634-472       &      15.2   &       337.4,      0.1   & 0.2 &      5.1   $\pm$ 0.5    & 0.3 &  1AGL J1639-4702  &  -  &  O \\ 
95     &      TeVJ1708-443       &      42.3   &      343.12,     -2.69  & 0.06&      13.9  $\pm$ 0.4    & 0.3 &      1AGL J1709-4428    &  3FGL J1709.7-4429 (E)    &  O \\
109    &      TeVJ1747-248       &      5.9    &         4.0,      1.7   & 0.3 &      2.1   $\pm$ 0.4    & 0.2 &    -  &  3FGL J1748.0-2447 (E)    &  O \\
112    &      TeVJ1801-233       &      15.0   &         6.6,      0.1   & 0.2 &      6.8   $\pm$ 0.5    & 0.4 &   1AGL J1801-2317  &  3FGL J1801.3-2326e (E)  &  E \\
113    &      TeVJ1804-216       & 	7.8    & 	 8.4,	   0.2	 & 0.3 & 	3.5	$\pm$	0.5	 & 0.2 & 	1AGLR J1805-2149	 &  3FGL J1805.6-2136e (E)  &  O \\%
116	 & 	TeVJ1813-178	 & 	4.8	 & 	13.0,	   0.4	 & 0.4 & 	2.1	$\pm$	0.5	 & 0.4 & 	1AGL J1815-1732	 &  -  &  IN \\
125	 & 	TeVJ1841-055	 & 	14.0	 & 	26.3,	   0.1	 & 0.2 & 	5.8	$\pm$	0.5	 & 0.6 & 	1AGLR J1839-0550	 &  3FGL J1840.9-0532e (E)	 &  O \\
128	 & 	TeVJ1848-017	 & 	4.6	 & 	30.8,	   0.1	 & 0.2 & 	1.8	$\pm$	0.4	 & 0.4 & 	-	 &  3FGL J1848.4-0141 (E)	 &  O \\
132	 & 	TeVJ1907+062	 & 	13.3	 & 	40.4,	  -1.0	 & 0.1 & 	4.4	$\pm$	0.4	 & 0.2 & 	1AGL J1908+0614	 &  3FGL J1907.9+0602 (E)	 &  O \\
133	 & 	TeVJ1911+090	 & 	7.8	 & 	43.3,	   0.0	 & 0.3 & 	2.4	$\pm$	0.3	 & 0.2 & 	-	 &  -  &  IN \\
135	 & 	TeVJ1923+141	 & 	7.1	 & 	49.2,	  -0.5	 & 0.3 & 	2.1	$\pm$	0.3	 & 0.2 & 	1AGL J1923+1404	 &  3FGL J1923.2+1408e (E)  &  IN \\%
142	 & 	TeVJ2019+368	 & 	25.5	 &     75.17,      0.25	 & 0.09& 	6.9	$\pm$	0.3	 & 0.2 & 	1AGLR J2021+3653	 &  3FGL J2021.1+3651 (E)  &  IN	\\  
144	 & 	TeVJ2032+415	 & 	8.0	 & 	80.3,	   1.2	 & 0.2 & 	2.3	$\pm$	0.3	 & 0.1 & 	1AGLR J2031+4130	 &  3FGL J2032.2+4126 (E)	 &  O \\
145	 & 	TeVJ2158-302	 & 	4.8	 & 	17.6,	  -52.0	 & 0.6 & 	1.4	$\pm$	0.3	 & 0.3 & 	-	 &  3FGLJ2158.8-3013 (P)	 &  IN \\
147	 & 	TeVJ2227+608	 & 	16.7	 &     106.7,	   3.0	 & 0.2 & 	3.3	$\pm$	0.2	 & 0.4 & 	1AGL J2231+6109	 &  3FGL J2225.8+6045 (E)	 &  E	\\
152	 & 	TeVJ2359-306	 & 	4.0	 & 	12.6,	  -78.0	 & 0.3 & 	3.3	$\pm$	1.0	 & 0.1 & 	-	 &  3FGL J2359.3-3038 (P)	 &  IN \\
\specialrule{0.11em}{0em}{0em}

38       &        TeVJ0835-455\footref{VelaX}  &        7.9      &        -, -    &        -        &  6.2 $\pm$ 0.8   &        -        &        -        &   3FGL J0833.1-4511e (E)              &  - \\
40       &        TeVJ0852-463\footref{VelaJr} &        4.0      &        -,-     &        -        &  1.5 $\pm$ 0.4   &        -        &        -        &   3FGL J0852.7-4631e (E)               &  - \\
64       &        TeVJ1325-430\footref{CenA}   &        4.4      &        -, -    &        -        &  0.8 $\pm$ 0.2   &        -        &        -          &   3FGL J1325.4-4301 (P)  &  - \\
89       &        TeVJ1640-465     &      12.3      &        -, -    &        -        &   5.1 $\pm$ 0.5   &        -        &        1AGL J1639-4702  &  3FGL J1640.4-4634c (E)   &  -  \\
96       &        TeVJ1713-382   & 	   6.5      & 	-, -    & 	-	 &  2.6 $\pm$ 0.4 & 	-	 & 	-	 &          -        &  - \\
98       &      TeVJ1714-385\footref{MLconfused}   &     7.0     &        -, -    &        -        &    2.8 $\pm$ 0.4   &        -        &        -        &  3FGL J1714.5-3832 (E)      &  - \\
99       &        TeVJ1718-374   & 	   6.2     & 	-, -	 & 	-	 &  2.6 $\pm$ 0.4   & 	-	 & 	-	 &  3FGL J1718.0-3726 (P)	 &  - \\
100	 & 	TeVJ1718-385	 & 	   6.9     & 	-, -	 & 	-	 &  2.7 $\pm$ 0.4   & 	-	 & 	-	 &  3FGL J1718.1-3825 (E)	 &  - \\ 
103	 & 	TeVJ1729-345	 &         8.0     & 	-, -	 & 	-	 &  3.6 $\pm$ 0.5   & 	-	 & 	-	 &  -  &  - \\
104	 & 	TeVJ1732-347	 & 	   7.3     & 	-, -	 & 	-	 &  3.1 $\pm$ 0.5   & 	-	 & 	-	 &  -  &  - \\ 
105      &        TeVJ1741-302\footref{gc}         &     4.3     & 	-, -	 & 	-	 & 2.1 $\pm$ 0.5 & 	-	 & 	-	 & 	-	 &  - \\
108	 & 	TeVJ1745-303	 & 	  11.2     & 	-, -	 & 	-	 &  5.5 $\pm$ 0.5   & 	-	 & 	1AGL J1746-3017	 &  3FGL J1745.1-3011 (E)	 &  - \\
118	 & 	TeVJ1825-137	 & 	   8.0     & 	-, -	 & 	-	 &  3.8 $\pm$ 0.5   & 	-	 & 	-	 &  3FGL J1824.5-1351e (E)	 &  - \\ 
119	 & 	TeVJ1826-148\footref{MLconfused}   &     7.7     & 	-, -	 & 	-	 &      3.3 $\pm$ 0.5    & 	-	 & 	-	 &  3FGL J1826.2-1450 (P)	 &  - \\ 
124      &        TeVJ1837-069   &       4.9       & 	        -, -	 & 	-	 &  2.1 $\pm$ 0.4   & 	-	 & 	-	 &  3FGL J1836.5-0655e (E)	 &  - \\ 
134      &        TeVJ1912+101   &       4.1       &  	-, -	 & 	-	 &  1.3 $\pm$ 0.3   & 	-	 & 	-	 & 	-	 &  - \\ 
139	 & 	TeVJ2001+438	 & 	 4.9       & 	-, -	 & 	-	 &  0.9 $\pm$ 0.2   & 	-	 & 	-	 &   3FGL J2001.1+4352 (P)	 &  - \\ 
146	 & 	TeVJ2202+422	 & 	 7.2       & 	-,-	 &      -        &  1.0 $\pm$ 0.2   &      -      &      -      &  3FGL J2202.7+4217 (P)	 &  - \\
150	 & 	TeVJ2323+588	 & 	 5.2       & 	-, -	 & 	-	 &  0.9 $\pm$ 0.2   & 	-	 & 	-	 &  3FGL J2323.4+5849 (P)	 &  - \\ 
\hline
\end{longtable}
}
\end{landscape}

%% file: TeV_paper_table3_3fgl.tex
\begin{landscape}
{\small
\LTcapwidth=\textwidth
\begin{longtable}{|c|c|c|c|c|c|c|}
\caption{Results of the spectral analysis of the most significant sources 
detected with AGILE in this analysis. The spectral indexes for all sources 
detected with a significance $\sqrt{(TS)_{3}}>5$ (see upper part of Table~\ref{tab:good_AGL})
and Galactic latitude $|b|<30^\circ$ are reported. The last column shows the Fermi
{\it power--law index} as given in the indicated reference; 
when the Fermi sources spectrum has been fitted with a {\it power--law} function
(i.e. the spectral form is {\tt PowerLaw}), the corresponding
error is also available and reported.}
\label{tab:spectrum_AGLgood} \\
\hline
ID & TeV Source & Type & AGILE power-law & AGILE & Fermi & 3FGL power-law \\
& & & spectral index\footnote{This analysis.} & Association & Association & spectral index\footnote{\citep{3FGL}. 
The errors on the Fermi power-law spectral indices in the 3FGL catalog are available only for three sources: 
3FGL J1633.0-4746e, 3FGL J1840.9-0532e, 3FGL J2225.8+6045. See also footnote \footref{powerlaw}.}\\
\hline
1   & TeVJ0006+727 & PWN     & 1.91 $\pm$ 0.05 & 1AGLR J0007+7307 & 3FGL J0007.0+7302 & 1.88 \\
12  & TeVJ0222+430 & IBL     & 2.04 $\pm$ 0.10 & 1AGLR J0222+4305 & 3FGL J0222.6+4301 & 1.94 \\
15  & TeVJ0240+612 & XRB     & 2.19 $\pm$ 0.04 & 1AGLR J0240+6115 & 3FGL J0240.5+6113 & 2.29 \\
19  & TeVJ0319+415 & FRI     & 1.80 $\pm$ 0.14 & 1AGLR J0321+4137 & 3FGL J0319.8+4130 & 2.08 \\
26  & TeVJ0534+220 & PWN     & 2.27 $\pm$ 0.03 & 1AGL J0535+2205  & 3FGL J0534.5+2201 & 2.23 \\
30  & TeVJ0616+225 & SNR     & 1.95 $\pm$ 0.07 & 1AGL J0617+2236  & 3FGL J0617.2+2234e & 1.98 \\
32  & TeVJ0632+173 & PWN     & 1.88 $\pm$ 0.02 & 1AGL J0634+1748  & 3FGL J0633.9+1746 & 1.87 \\
36  & TeVJ0721+713 & LBL     & 1.87 $\pm$ 0.07 & 1AGLR J0723+7121 & 3FGL J0721.9+7120 & 2.04 \\
45  & TeVJ1018-589 & BIN     & 2.07 $\pm$ 0.11 & 1AGLR J1018-5852 & $ \begin{cases} \textrm{3FGL J1016.3-5858} \\ \textrm{3FGL J1018.9-5856} \end{cases} $ & $ \begin{aligned} 2.35 \\ 2.30 \end{aligned} $ \\
66  & TeVJ1418-609 & PWN     & 2.00 $\pm$ 0.07 & 1AGLR J1417-6108 & 3FGL J1418.6-6058 & 2.27 \\
75  & TeVJ1459-608 & PWN     & 2.16 $\pm$ 0.12 & - & $ \begin{cases} \textrm{3FGL J1456.7-6046} \\  \textrm{3FGL J1459.4-6053} \end{cases} $ &  $ \begin{aligned} 2.37 \\ 2.37 \end{aligned} $ \\
87  & TeVJ1632-478 & UNID    & 2.05 $\pm$ 0.14 & - & 3FGL J1633.0-4746e & 2.11 $\pm$ 0.02 \\
88  & TeVJ1634-472 & UNID    & 2.58 $\pm$ 0.08 & 1AGL J1639-4702 & - & - \\
95  & TeVJ1708-443 & PWN     & 2.02 $\pm$ 0.02 & 1AGL J1709-4428 & 3FGL J1709.7-4429 & 2.02 \\
109 & TeVJ1747-248 & GC      & 1.99 $\pm$ 0.17 & - & 3FGL J1748.0-2447 & 2.26 \\
112 & TeVJ1801-233 & SNR/MC  & 2.07 $\pm$ 0.07 & 1AGL J1801-2317  & 3FGL J1801.3-2326e & 2.15 \\
113 & TeVJ1804-216 & UNID    & 2.13 $\pm$ 0.12 & 1AGLR J1805-2149 & 3FGL J1805.6-2136e & 2.07 \\
125 & TeVJ1841-055 & UNID    & 2.41 $\pm$ 0.08 & 1AGLR J1839-0550 & 3FGL J1840.9-0532e & 1.81 $\pm 0.04$ \\
132 & TeVJ1907+062 & UNID    & 2.36 $\pm$ 0.07 & 1AGL J1908+0614  & 3FGL J1907.9+0602 & 2.24 \\
133 & TeVJ1911+090 & SNR/MC  & 2.21 $\pm$ 0.12 & - & - & - \\
135 & TeVJ1923+141 & SNR/MC  & 2.23 $\pm$ 0.12 & 1AGL J1923+1404  & 3FGL J1923.2+1408e & 2.15 \\
142 & TeVJ2019+368 & PWN     & 1.98 $\pm$ 0.04 & 1AGLR J2021+3653 & 3FGL J2021.1+3651 & 2.17 \\
144 & TeVJ2032+415 & UNID    & 2.05 $\pm$ 0.10 & 1AGLR J2031+4130 & 3FGL J2032.2+4126 & 2.08 \\
147 & TeVJ2227+608 & SNR     & 2.30 $\pm$ 0.06 & 1AGL J2231+6109  & 3FGL J2225.8+6045 & 1.95 $\pm$ 0.20 \\
\hline
\end{longtable}
}
\end{landscape}

%% file: TeV_paper_table4.new.tex
\begin{table}[ht]
\caption{Statistics about the sources detected by AGILE in this work, grouped by 
source type classification. Enclosed in parenthesis, the percentage of detected sources with respect to the total class sample.}
\begin{center}
\begin{tabular}{|l|c||c|l|}
\hline
{\it Source type} & {\it Detected / Total} & {\it Source class} & {\it Detected / Total} \\
\hline
\multirow{7}{*}{EXTRAGALACTIC} & \multirow{7}{*}{13 / 66 (20\%)} & Blazar & 0 / 1 (0\%)\\
 & & HBL & 5 / 44 (11\%) \\
 & & IBL  & 2 / 7 (29\%)\\
 & & LBL  & 2 / 3 (67\%)\\
 & & FSRQ & 2 / 5 (40\%)\\
 & & Sbs  & 0 / 2 (0\%) \\
 & & Superbubble & 0 / 1 (0\%) \\
 & & FRI  & 2 / 3 (67\%)\\
\hline
 \multirow{7}{*}{GALACTIC} & \multirow{7}{*}{30 / 58 (52\%)} & PWN & 11 / 28 (39\%)\\
 & & SNR     & 7 / 11 (64\%)  \\ 
 & & PWN/SNR & 2 / 2 (100\%) \\
 & & SNR/MC & 5 / 8 (63\%) \\
 & & BIN/XRB & 3 / 5 (60\%)  \\
 & & GC      & 1 / 1 (100\%) \\
 & & WR      & 1 / 3 (33\%)\\
\hline
\multirow{3}{*}{UNIDENTIFIED} & \multirow{3}{*}{9 / 28 (32\%)} & & \\
 & & - & - \\
 & & & \\
\hline
\end{tabular}
\end{center}
\label{tab:statistics}
\end{table}